\documentclass[final,3p,times,11pt]{elsarticle}

\usepackage[USenglish]{babel}
\usepackage{amsmath,amssymb,amsthm, mathrsfs}
\usepackage{mathtools}
\usepackage{graphicx}
\usepackage{stmaryrd}
\usepackage[dvipsnames]{xcolor}
\usepackage{cancel}
\usepackage{ulem}
\usepackage{showlabels}
\usepackage[linesnumbered,ruled,vlined]{algorithm2e}
\usepackage{hyperref}
\hypersetup{
    colorlinks=true,       
}

\usepackage{wrapfig}  
\usepackage{marginnote}
\usepackage{multirow}

\newtheorem{theorem}{Theorem}
\AfterEndEnvironment{theorem}{\noindent\ignorespaces}

\journal{}
\makeatletter
\def\ps@pprintTitle{%
 \let\@oddhead\@empty
 \let\@evenhead\@empty
 \def\@oddfoot{}%
 \let\@evenfoot\@oddfoot}
\makeatother

\begin{document}
\begin{frontmatter}

\title{Multilevel Monte Carlo methods for the Grad-Shafranov free boundary problem}



\author[umdcs]{Howard C.\ Elman}
\ead{helman@umd.edu}
\address[umdcs]{Department of Computer Science and Institute for Advanced Computer
Studies, University of Maryland, College Park.}
\author[umdm]{Jiaxing Liang}
\ead{jliang18@umd.edu}
\address[umdm]{Applied Mathematics \& Statistics and Scientific Computing Program, University of Maryland, College Park.}
\author[UA]{Tonatiuh S\'anchez-Vizuet}
\ead{tonatiuh@math.arizona.edu}
\address[UA]{Department of Mathematics, The University of Arizona.}
\begin{abstract}
The equilibrium configuration of a plasma in an axially symmetric reactor is described mathematically by a free boundary problem associated with the celebrated Grad-Shafranov equation. The presence of uncertainty in the model parameters introduces the need to quantify the variability in the predictions. This is often done by computing a large number of model solutions on a computational grid for an ensemble of parameter values and then obtaining estimates for the statistical properties of solutions. In this study, we explore the savings that can be obtained using multilevel Monte Carlo methods, which reduce costs
by performing the bulk of the computations on a sequence of spatial grids that are coarser than the one that would typically be used for a simple Monte Carlo simulation. We examine this approach using both a set of uniformly refined grids and a set of adaptively refined grids guided by a discrete error estimator. Numerical experiments show that multilevel methods dramatically reduce the cost of simulation, with cost reductions
typically on the order of 60 or more and possibly as large as 200. Adaptive gridding results in more accurate computation of geometric quantities such as $x$-points associated with the model.
\end{abstract}

\end{frontmatter}


\section{Introduction}\label{sec:intro}
Monte Carlo (MC) techniques are one of the most common strategies for dealing with the quantitative assessment of the accuracy of numerical simulations of physical models with uncertainties. The idea behind these methods is to obtain a large number of samples (typically by numerically solving the associated model) for random realizations of the uncertain parameters, and use these data to gather statistical information about the quantity of interest.  However, when the model involves the solution of partial differential equations, the computational effort related to the collection of the data points required can easily become unmanageable. To overcome this difficulty methods like polynomial chaos expansions \cite{Xiu:2010}, stochastic Galerkin \cite{GhSp:1993}, and stochastic collocation \cite{BaNoTe:2007} have been used to handle uncertainties associated with a small number of parameters. These techniques, however, often require the development of specialized numerical solvers or rely on the smooth dependence of the model with respect to the parameter values. The multilevel Monte Carlo (MLMC) method was developed \cite{ClGiScTe:2011, Gi:2015,NoTe:2015,TeScGiUl:2013} as an efficient alternative that does not require additional smoothness assumptions and can take advantage of an existing numerical solver. Given a target numerical grid (i.e. a grid whose resolution is considered sufficiently fine) MLMC improves the efficiency of the sampling step by offsetting the bulk of the numerical computations to a sequence of \textit{coarser} grids where the numerical solution is cheaper. 

In the particular context of free-boundary Grad-Shafranov computations subject to parameter uncertainty, the authors have shown that the computational cost can be reduced manyfold by employing a strategy based on stochastic collocation \cite{ElLiSa:2022}; however, due to the latent possibility of plasma-wall contacts, the smoothness of the mapping between coil currents and equilibria cannot be guaranteed. In this paper, our goal is to overcome this difficulty by approximating the expectation of the equilibrium configuration using a multilevel Monte Carlo Finite-Element (MLMC-FE) approach. We will consider two MLMC-FE algorithms: a classical strategy based on uniformly refined meshes and a variation based on meshes refined adaptively using an \textit{a posteriori} error estimator. As we shall see, both of these approaches greatly reduce computational costs, with the adaptive strategy being somewhat more effective in computing the approximation of the expectation of geometric properties of the equilibrium configuration.  

An outline of the paper is as follows. In Section \ref{sec:Grad-Shafranov}, we briefly recall the Grad-Shafranov free boundary problem. Section \ref{sec:MLMC-FE} is devoted to the introduction of the Monte Carlo and multilevel Monte Carlo Finite-Element methods. The section concludes with the introduction of an algorithm to compute the optimal number of samples required at each discretization level. In Section \ref{sec:Num-Exp} we present numerical experiments comparing the effectiveness of these strategies with a Monte Carlo strategy based on a single mesh. Concluding remarks are presented in the final Section \ref{sec:Conclusion}. For completeness, the technical mathematical, and algorithmic aspects of the problem and the methods discussed are included as an appendix.
\section{The Grad-Shafranov free boundary problem}\label{sec:Grad-Shafranov}
\subsection{The deterministic problem}

\begin{wrapfigure}[17]{r}{0.3\textwidth}
\begin{center}\includegraphics[width=0.65\linewidth]{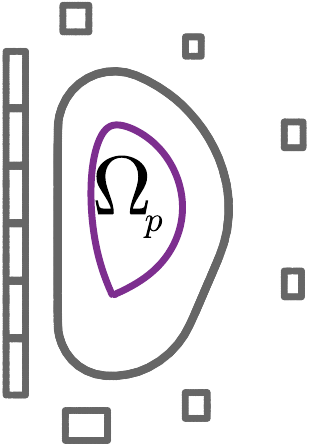}
\end{center}
\caption{The plasma confinement region $\Omega_p$ is enclosed by the violet line. The rectangles represent the external coils $C_k$; the grey curved line represents the exterior wall of the vacuum chamber. }\label{fig:Geometry}
\end{wrapfigure}

In a cylindrically symmetric magnetic confinement device, with coordinates $(r,z,\varphi)$, the mathematical expression of the equilibrium condition between the hydrostatic and magnetic forces acting on the plasma results in the celebrated Grad-Shafranov equation \cite{GrRu:1958,LuSc:1957,Shafranov:1958}. This nonlinear elliptic equation relates the \textit{poloidal flux function} $\psi(r,z)$ to the hydrostatic pressure $p(\psi)$ and the \textit{the toroidal field} $g(\psi)$ (both of which are assumed to be functions of $\psi$ only), and the currents $I_k$ going through external coils with cross-sectional area $S_k$. Posed in free space, the equation takes the form
\begin{subequations}\label{eq:FreeBoundary}
\begin{equation}\label{eq:FreeBoundarya}
 -\nabla\,\cdot\,\left(\frac{1}{\mu r}\nabla\psi\right) = \left\{ \begin{array}{ll}
r\frac{d}{d\psi} p(\psi) + \frac{1}{2\,\mu r} \frac{d}{d\psi} g^2(\psi) & \text{ in } \Omega_p(\psi) \\
I_k/S_k & \text{ in } \Omega_{C_k} \\
0 & \text{ elsewhere. } 
\end{array}\right.
\end{equation}
Above, $\mu$ is the magnetic permeability, $\Omega_{C_k}$ denotes the area occupied by the $k$-th external coil, $\Omega_p$ is the plasma confinement region which \textit{is not known a priori} and must be determined as a problem unknown, making this a \textit{free boundary problem}. A schematic of a  cross-section, for $r>0$, of a tokamak is depicted in Figure \ref{fig:Geometry}. The confinement region $\Omega_p$ is characterized as the largest region that contains the magnetic axis (defined as the point where $\psi$ has a global maximum) and that is bounded by a closed level set $\psi=constant$. The solution to this free boundary problem is ubiquitous in nuclear fusion and several computational codes have been developed over the years (see for instance \cite{FaHe:2017,GoLeNe:2006,CEDRES,HeRa:2017,Hofmann1988,Johnson1979} and references therein). 

A common choice in the literature, first proposed in \cite{LuBr:1982}, for the free functions $p(\psi)$ and $g(\psi)$ in the right hand side of \eqref{eq:FreeBoundarya} is
\begin{equation}\label{eq:source}
\frac{d}{d\psi}p(\psi) = j_0\frac{\beta}{r_0}\left(1-\psi_N^{\alpha_1}\right)^{\alpha_2},  \qquad \qquad
\frac{1}{2}\frac{d}{d\psi}g^2(\psi) = j_0\mu_0r_0(1-\beta)\left(1-\psi_N^{\alpha_1}\right)^{\alpha_2},
\end{equation}
\end{subequations}

\!\!\!\!\!\!\!\!\!\!where $r_0$ represents the outer radius of the vacuum chamber and $\psi_{N}\in[0,1]$ is a normalization of the flux $\psi$ such that $\psi_N = 1$ on the plasma boundary $\partial\Omega_p$ and $\psi_N = 0$ at the magnetic axis. The parameters $\alpha_1$ and $\alpha_2$ control the behavior of $\psi$ around the magnetic axis, and $\beta$ measures the ratio between the hydrostatic pressure and the magnetic pressure in the plasma, and $j_0$ is a normalization factor. 
%
\subsection{Incorporating uncertainty}
%
In this article, we will consider that the uncertainty in the model \eqref{eq:FreeBoundarya} is concentrated in the values of the currents $I_i$ going through the external coils. As a result, the function $\psi$ (and all quantities derived from it) are random variables. Obtaining a full description of their probability density functions might not be possible, but an approximate picture can be obtained by exploring the parameter space and computing sample approximations of its expectation and variance. We will model the array of currents as a $d$-dimensional random variable $\boldsymbol \omega :=(\omega_1,\ldots,\omega_d)$, where $d$ is the number of confinement coils in the reactor, and the $k$-th component of $\boldsymbol \omega$ is the current going through the $k$-th coil. We will consider that $\boldsymbol\omega$ is uniformly distributed around a baseline vector $\boldsymbol I = (I_1,\ldots,I_d)$ corresponding to the desired current values in a deterministic model. We will often refer to $\boldsymbol I$ as either the \textit{reference} or \textit{unperturbed} currents. Letting $\tau>0$ denote the size of the possible perturbation in the current values (relative to the components of $\boldsymbol I$), the vector $\boldsymbol \omega$ is then uniformly distributed over the $d$-dimensional parameter space
\begin{equation}
\label{eq:ParameterSpace}
    W := \prod_{k=1}^{d}\big[I_k-\tau|I_k|,I_k+\tau|I_k|\big].
\end{equation}

Since coils are independent of each other, the stochastic random currents $\left\{\omega_k\right\}_{k=1}^{d}$ are uncorrelated and the joint density function of $\boldsymbol{\omega}$ is given by $\pi \left(\boldsymbol{\omega}\right)=\prod_{k=1}^{d} \pi_k\left(\omega_{k}\right)=\prod_{k=1}^{d} \frac{1}{2\tau |I_k|}$.  The equilibrium configuration determined by the solution to \eqref{eq:FreeBoundarya} is then the random variable $\psi(x,y,\boldsymbol{\omega})$; we will be primarily interested in efficiently constructing an approximation to its expected value
 \begin{equation}
 \label{eq:QoI}
      \mathbb{E}\left[\psi(x,y,\boldsymbol \omega)\right]=\int_W \psi(x,y,\boldsymbol{\omega})\pi(\boldsymbol\omega)d\boldsymbol{\omega}, 
 \end{equation}
as well as those of some derived quantities such as the plasma boundary, the location of the x-points, etc.
\section{Monte Carlo and Multilevel Monte Carlo Finite-Element Methods}
\label{sec:MLMC-FE}

We now turn our attention to the numerical approximation of the expected value \eqref{eq:QoI}. Since the location and shape of the plasma boundary depend on the values of the coil currents, variations of these values could lead to contacts between the plasma and the wall or even loss of confinement. This fact translates into a possible non-smoothness of the mapping between coil currents and the solutions of \eqref{eq:FreeBoundarya} which may then cause techniques such as stochastic collocation to underperform. Moreover, the computational effort associated with cubature methods scales exponentially with the dimension of the parameter space, seriously limiting their feasibility for estimating  \eqref{eq:QoI}. This leads to the use of Monte Carlo methods, which are agnostic to both the smoothness of the mapping and the dimensionality of the problem \cite{MeUl:1949}, although they have a slow convergence rate (1/2) that tends to make them costly. This will be addressed through the use of a multi-level approach.

\subsection{Monte Carlo Finite-Element method}

We will describe the method in terms of a generic solution, $u$, to a PDE involving stochastic parameters and its finite element approximation $u_h$, where $h$ is the mesh parameter of the discretization. Let $\{\boldsymbol{\omega}^{(i)}\}_{1\le i \le N}$ be a set of $N$ realizations of the random variable $\boldsymbol\omega$  giving rise to a sample of $N$ realizations $u^{(i)} = u(\boldsymbol \omega^{(i)})$ and $u_h^{(i)}=u_h(\boldsymbol{\omega}^{(i)})$ of the exact solution and its finite element discretization. We will assume that all these functions belong to a functional space $Z$ endowed with a norm $\|\cdot\|_Z$ (see the Appendix \ref{sec:Space_Norm_VariationalFormulation} for details),
and we will consider the standard FEM error estimate $\|u^{(i)}-u^{(i)}_h\|_Z\leq C^{(i)}h^p$,
where $p$ is the order of the FEM discretization and the constant $C^{(i)}$ depends only on the problem geometry and the particular values of $\boldsymbol\omega^{(i)}$. 

The Monte Carlo Finite-Element (MC-FE) estimator $A_{\text{MC}}(u_h)$ for $\mathbb{E}(u)$ is defined as the sample mean
\begin{equation}
    A_{\text{MC}}(u_h):= \frac{1}{N}\sum_{i=1}^{N}u_h^{(i)}.
\end{equation}
This estimator is easily shown to be unbiased and to satisfy $\mathbb{E}(A_{\text{MC}}) = \mathbb{E}(u_h)$.  A quantity that serves as a foundation for examining the spatial and statistical accuracy of the MC-FE estimator is the \textit{mean squared error} (MSE) defined as
 \[
\mathcal{E}_{A_\text{MC}}^2:=\mathbb E\left[\left\Vert\mathbb{E}(u)-A_{\text{MC}}(u_h) \right\Vert_{Z}^2\right].
 \]
It can be shown (see, for instance \cite[Theorem~4.3]{BaScZo:2011}) that, for linear problems, the Monte Carlo estimator accurately approximates the expected value in the sense that
$\mathcal{E}_{A_\text{MC}}^2 \le K(N^{-1/2}+h^p)^2$,
 where the constant $K>0$ depends on the problem geometry and the expected values of the problem data.
 The MSE can be decomposed into terms related to the bias and variance as
\[
\mathcal{E}_{A_\text{MC}}^2= \left\Vert\mathbb{E}(u)-\mathbb{E}(u_h)\right\Vert^2_Z+\mathbb{E}\left[\left\Vert\mathbb{E}(u_h)-A_{\text{MC}}(u_h)\right\Vert_Z^2\right]
    =\left\Vert\mathbb{E}(u)-\mathbb{E}(u_h)\right\Vert^2_Z + \frac{\mathbb{V}(u_h)}{N} =\mathcal{E}_{\text{Bias}}^2 + \mathcal{E}_{\text{Stat}}^2   ,
\]
where $\mathbb{V}(u) := \mathbb{E}[\left\Vert u - \mathbb{E}(u)\right\Vert_Z^2]$ and $\mathbb{V}(A_{\text{MC}}(u)) = \mathbb{V}(u)/N$. The last two terms in the
expression above implicitly define the discretization error $\mathcal{E}_{\text{Bias}}$ and the sampling (or statistical) error $\mathcal{E}_{\text{Stat}}$ respectively.

If $\mathbb{E}(u)\neq 0$, the mean squared error can be expressed as a percentage through normalization by the factor $\left\Vert\mathbb{E}(u) \right\Vert_{Z}^2$, leading to the \textit{normalized mean squared error} $\widehat{\mathcal{E}}_{A_\text{MC}}^2$. Since the exact random variable $u$ is not available, we will approximate the relative mean squared error (nMSE) by
 
\begin{equation}
 \label{eq:MC_nMSE}    	\widehat{\mathcal{E}}_{A_\text{MC}}^2\approx \frac{\left\Vert\mathbb{E}(u)-\mathbb{E}(u_h)\right\Vert^2_Z}{\left\Vert\mathbb{E}(u_h) \right\Vert_{Z}^2} + \frac{\mathbb{V}(u_h)}{N\left\Vert\mathbb{E}(u_h) \right\Vert_{Z}^2} = \widehat{\mathcal{E}}_{\text{Bias}}^2 + \widehat{\mathcal{E}}_{\text{Stat}}^2,
\end{equation}
 where $\widehat{\mathcal{E}}_{\text{Bias}}$ and  $\widehat{\mathcal{E}}_{\text{Stat}}$ are relative analogues to the discretization and statistical errors defined above. If the number of grid points for the FEM discretization is $M$ then, in two dimensions, it is standard to assume that $\widehat{\mathcal{E}}_{\text{Bias}} = \mathcal{O}(M^{-p/2})$. Given a target tolerance $\epsilon$, the contribution of the statistical and discretization errors towards the total nMSE can be controlled by requiring that  
\begin{equation}
\label{eq:nMSE_MC_ErrSplitting}
\widehat{\mathcal{E}}_{\text{Bias}}^2 = \mathcal{O}(M^{-p})\le (1-\theta)\epsilon^2,\qquad \widehat{\mathcal{E}}_{\text{Stat}}^2 = \frac{V_h}{N} \le \theta \epsilon^2, 
\end{equation}
where $\theta\in(0,1)$ is known as the \textit{splitting parameter}, and $V_h:=\mathbb{V}\left(u_h\right)/ \left\Vert\mathbb{E}(u_h) \right\Vert_{Z}^2$. This in turn allows us to estimate the sample size $N$ and the number of grid points $M$ required to achieve the desired tolerance as 
\begin{equation}
\label{eq:MC_SampleSizeEstimate}
M\ge \left((1-\theta)\epsilon^2\right)^{-\frac 1 {p}},\qquad N =  \Biggl\lceil \frac{V_h}{\theta \epsilon^2} \Biggr\rceil.
\end{equation}

Assuming the average cost to obtain one sample (i.e. to solve \eqref{eq:FreeBoundarya} for one  
value of the coil-currents) is $C =\mathcal{O}(M^c$) for some $c>0$, the total computational cost of the MC-FE estimator can be estimated as 
\begin{equation}
\label{eq:MC_Cost}
C(A_{\text{MC}})=\mathcal{O}(NM^c) =\mathcal{O}\left(\epsilon^{-2-\frac{2c}{p}}\right).
\end{equation}

%
\subsection{Multilevel Monte Carlo Finite-Element method}
%

MLMC reduces the computational cost associated with 
sampling---which in our case 
involves the numerical solution of a non-linear PDE in a target computational mesh---by approximating the expectation of the quantity of interest on the finest mesh by a sequence of control variates on a set of coarse grids that are cheaper to compute \cite{FaDoKe:2017}. 
Using the linearity of expectation, the MLMC estimator expresses the quantity of interest on the finest spatial grid, $\mathbb E\left(u_h\right)$, by a telescoping sum involving the numerical approximations of $u$ on coarser grids. Consequently, MLMC's workload is shifted from the fine grid to coarser grids, making it more efficient than MC \cite{ClGiScTe:2011}. Cf. e.g. \cite{Law-Cerfon-Pehersdorfer} for alternative ways to reduce the costs of MC methods.

To construct meshes that are easy to describe for both uniform and adaptive mesh refinement, we will characterize them using the number of grid points rather than the mesh size. We will refer to $\ell = 0, \dots, L$ as the \textit{level} of a mesh $\mathcal{T}_\ell$ containing $\{M_\ell\}$ grid points. We will then consider a sequence of meshes $\mathcal T_0,\ldots,\mathcal T_L$ with increasing resolution so that $\{M_\ell\}_{0\le \ell \le L}$ defines an increasing sequence and $\mathcal T_L$ is the finest mesh, and we will denote by $u_\ell$ the approximation of $u$ on the mesh $\mathcal{T}_\ell$. 
	
The expectation of the function $u$ can be approximated by the expectation of the finest approximation $u_L$. 
Since $u_L = u_0 + (u_1-u_{0}) + (u_2-u_1) + \cdots + (u_L - u_{L-1})$, this approximation is given by
 \begin{equation}\label{eq:MLMCapproxMean}
\mathbb{E}(u)\approx\mathbb{E}(u_L)=\mathbb{E}(u_0)+\sum_{\ell=1}^L\mathbb{E}(u_\ell-u_{\ell-1})= \sum_{\ell=0}^L\mathbb{E}(Y_\ell),
\end{equation}
a telescoping sum, where each of the terms 
\begin{equation}\label{eq:Y_l}
Y_0:=u_0 \quad  \text{ and } \quad Y_\ell := u_\ell - u_{\ell-1} \quad (\text{for } \ell\ge 1)
\end{equation}
can be regarded as a correction of the coarsest approximation $u_0$. If each term $\mathbb{E}(Y_\ell)$ is estimated by gathering $N_\ell$ samples at level $\ell$ and computing the sample expectations
\[
\mathbb{E}(Y_0) \approx \widehat{Y}_0:=\frac{1}{N_0}\sum_{i=1}^{N_0}u_0^{(i)}, \qquad \mathbb{E}(Y_\ell)\approx \widehat{Y}_\ell:=\frac{1}{N_\ell}\sum_{i=1}^{N_\ell}\left(u_\ell^{(i)}-u_{\ell-1}^{(i)}\right) \quad (\text{ for }\ell \ge 1),
\]
then the MLMC-FE estimator at level $L$ will be unbiased and can be written as
\begin{equation}
\label{eq:MLMC_estimator}
    A_{\text{MLMC}}(u_L) := \sum_{\ell=0}^L \widehat{Y}_\ell=\frac{1}{N_0}\sum_{i=1}^{N_0}u_0^{(i)} + \sum_{\ell = 1}^L \frac{1}{N_\ell}\sum_{i=1}^{N_\ell}\left(u_\ell^{(i)} - u_{\ell-1}^{(i)}\right).
\end{equation}
Recalling that $\mathbb{E}(\widehat{Y}_\ell) = \mathbb{E}(Y_\ell)$ and $\mathbb{V}(\widehat{Y}_\ell) = \mathbb{V}(Y_\ell)/N_\ell$ we conclude that, for the MLMC-FE estimator, it follows that $\mathbb{E}\left(A_{\text{MLMC}}\right) = \sum_{\ell=0}^L \mathbb{E}(\widehat{Y}_\ell)=\mathbb{E}(u_L)$ and $\mathbb{V}\left(A_{\text{MLMC}}\right) =\sum_{\ell=0}^L\mathbb{V}(\widehat{Y}_\ell )= \sum_{\ell=0}^L \mathbb{V}(Y_\ell)/N_\ell$.

As for MC-FE, the mean squared error 
$\mathcal{E}_{A_\text{MLMC}}^2$ 
can be split into contributions from bias and variance as
\begin{equation*}
\mathcal{E}_{A_\text{MLMC}}^2=\mathbb{E} \left[\left\Vert\mathbb{E}(u)-A_{\text{MLMC}}(u_L) \right\Vert_{Z}^2\right]=\left\Vert\mathbb{E}(u)-\mathbb{E}(u_L)\right\Vert^2_Z + \sum_{\ell = 0}^L\frac{\mathbb{V}\left(Y_\ell\right)}{N_\ell} = \mathcal{E}_{\text{Bias}}^2 + \mathcal{E}_{\text{Stat}}^2.
\end{equation*}
Similarly, 
using $\mathbb E(u)\approx \mathbb E(u_L)$, 
the normalized mean squared error $ \widehat{\mathcal{E}}_{A_\text{MLMC}}^2$ can be approximated by
\begin{equation}
\label{eq:MLMC_nMSE}
    \widehat{\mathcal{E}}_{A_\text{MLMC}}^2\approx \frac{\left\Vert\mathbb{E}(u)-\mathbb{E}(u_L)\right\Vert^2_Z}{\left\Vert\mathbb{E}(u_L) \right\Vert_{Z}^2}+\sum_{\ell = 0}^L\frac{V_\ell}{N_\ell} = \widehat{\mathcal{E}}_{\text{Bias}}^2 + \widehat{\mathcal{E}}_{\text{Stat}}^2,
\end{equation}
where
\begin{equation}\label{eq:NormalizedVariance}
V_\ell:=\mathbb{V}\left(Y_\ell\right)/ \left\Vert\mathbb{E}(u_L) \right\Vert_{Z}^2.
\end{equation}
As above, the parameter $\theta\in(0,1)$ can be used to split the contributions of the 
two components of the error by requiring 
$\widehat{\mathcal{E}}_{\text{Bias}}^2 \le (1-\theta)\epsilon^2$ and $\widehat{\mathcal{E}}_{\text{Stat}}^2 \le \theta\epsilon^2$ where $\epsilon$ is a predetermined tolerance such that $\widehat{\mathcal{E}}_{A_\text{MLMC}}^2\le \epsilon^2$. 

Note that each of the terms $ Y_\ell^{(i)} := u_\ell^{(i)}-u_{\ell-1}^{(i)}$ appearing in \eqref{eq:MLMC_estimator} requires the approximation of $u^{(i)}$ on adjacent refinement levels \textit{using the same value of the parameter} $\boldsymbol \omega^{(i)}$. However,  
for FEM discretization, the numerical implementation of this term does not require the solution of the PDE on the two grids $\mathcal{T}_{\ell}$ and $\mathcal{T}_{\ell-1}$. The realization on the coarse grid can be obtained from the one on the fine grid by either Galerkin projection or interpolation. Projection is the more accurate choice, but it requires the solution of a system involving the mass matrix for each realization. Thus we prefer interpolation, for which costs are minimal and which is of second-order accuracy in our application. To avoid introducing correlation across discretization levels, none of the samples involved in the computation of $Y_\ell$ is reused for the finer level $\ell+1$.  
That is, sampling is done such that, for $n\neq m$, the estimates $Y_n$ and $Y_m$ are uncorrelated. However, for \textit{any particular} $Y_\ell$ the strong correlation between  
$u_\ell^{(i)}$ and $u_{\ell-1}^{(i)}$ makes the variance of the correction terms much smaller than the variance of the approximation $u_L$ in the finest mesh, further improving the statistical approximation. 

To quantify the computational effort of the MLMC-FE estimator, let $M_{\ell,i}$ be the number of grid points for the $i$-th sample on mesh level $\ell$. We will assume that the computational cost to obtain one sample of $u_\ell^{(i)}$ is $C_{\ell,i} :=C(u_\ell^{(i)}) = \mathcal{O}(M_{\ell,i}^{c})$, where the exponent $c>0$ depends on the solver, and will denote the cost of computing the correction term $Y_\ell^{(i)}$ by $C_{\ell,i}:=C(Y_\ell^{(i)}) = \mathcal{O}(M_{\ell,i}^{c})$ for $\ell\ge 0$ and $M_{-1,i}=0$. For a nonlinear problem like the one at hand, the particular realization $\boldsymbol \omega^{(i)}$ will 
influence the cost. 
We will consider the \textit{average} cost to be of the form
\begin{equation}\label{eq:AverageCost}
C_\ell = \mathcal{O}(M_{\ell}^{c})
\end{equation}
and use this to estimate the total cost as
$C(A_{\text{MLMC}}) = \sum_{\ell=0}^L N_\ell C_{\ell}=\sum_{\ell=0}^L N_\ell\cdot \mathcal{O}\left(M_\ell^{c}\right).$
Using the method of Lagrange multipliers, it is shown in \cite{Gi:2015}
that this total cost, a function of $N_{\ell}$, can be 
minimized subject to the constraint 
$\widehat{\mathcal{E}}_{\text{Stat}}^2\le \theta \epsilon^2$
by the choice 
\begin{equation}
\label{eq:MLMC_SampleSize}
		N_{\ell} =  \Biggl\lceil \frac{1}{\theta \epsilon^2} \sqrt{\frac{V_\ell}{C_\ell}}\sum_{k = 0}^{L}\sqrt{V_k C_k}\Biggr\rceil.
\end{equation}
With this expression for $N_\ell$, the optimal total cost for the MLMC-FE estimator is 
\begin{equation}
\label{eq:MLMC_Cost}
    C(A_{\text{MLMC}}) \le\frac{1}{\theta \epsilon^2}\left(\sum_{\ell=0}^L \sqrt{V_\ell C_\ell}\right)^2 + \sum_{\ell=0}^L C_\ell.
\end{equation}
This analysis of computational cost is established in full 
generality in \cite[Theorem 1]{Gi:2015}.

The formula \eqref{eq:MLMC_SampleSize} suggests an iterative procedure for the approximation of $\mathbb E(u)$. Starting from a computational mesh $\mathcal T_0$, gather an initial number $\widetilde M_0$ of samples $u_0^{(i)}$ and estimate $\widehat{\mathcal E}_{\text{Bias}}$, $\widehat{\mathcal E}_{\text{Stat}}$, and $V_0$. If $\widehat{\mathcal E}_{\text{Stat}}$ is larger than the prescribed tolerance, use \eqref{eq:MLMC_SampleSize} to update $\widetilde M_0$ and gather additional samples; if $\widehat{\mathcal E}_{\text{Bias}}$ is above the prescribed tolerance, then add an additional level of spatial refinement. The process continues adding discretization levels and collecting additional and/or samples until both $\widehat{\mathcal E}_{\text{Bias}}$ and $\widehat{\mathcal E}_{\text{Stat}}$ fall below the required tolerance, at which point $\mathbb E(u)$ is approximated using equation \eqref{eq:MLMCapproxMean}. This simple algorithm, however, presents one challenge: the term $V_\ell$ in equation \eqref{eq:MLMC_SampleSize} requires the computation of the term  
\begin{equation}\label{eq:MeanVarUpdate_Var}
\mathbb{V}(Y_\ell) = \frac{1}{N_\ell-1}\left(\sum_{i=1}^{N_\ell}\left\Vert Y_\ell^{(i)}\right\Vert_{Z}^2-\frac{1}{N_\ell}\left\Vert\sum_{i=1}^{N_\ell}Y_\ell^{(i)}\right\Vert_{Z}^2\right).
\end{equation}
However, the estimated sample sizes $\{N_\ell\}$ are available only for pre-existing discretization levels, hence whenever an additional mesh refinement is needed, $N_{L+1}$ cannot be approximated by  \eqref{eq:MLMC_SampleSize} as this formula uses $V_{L+1}$ to compute $N_{L+1}$. This inconvenience can be overcome by noting that 
\begin{equation}\label{eq:EstimateV_l}
\mathbb{V}(u-u_\ell) = \mathbb{E}\left[\left\Vert u-u_\ell\right\Vert_Z^2\right] - \left\Vert \mathbb{E}\left(u-u_\ell\right)\right\Vert_Z^2 \le \mathbb{E}\left[\left\Vert u-u_\ell\right\Vert_Z^2\right].
\end{equation}
Hence, the variance $\mathbb{V}(u-u_\ell)$ can be estimated by the expectation of the squared discretization error $\left\Vert u-u_\ell\right\Vert_Z^2$. For a uniformly refined grid, we can resort to a standard \textit{a priori} error estimate and assume that $\mathbb{E}[\left\Vert u-u_\ell\right\Vert_Z^2]=\mathcal{O}( M_\ell^{-b_1})$ for some $b_1>0$.  It then follows that $V_\ell=\mathcal{O}(M_\ell^{-b_1})$ and following \cite{MoAp:2018} we can then approximate   $V_{L+1}$ in terms of the known variance $V_{L}$ by
\begin{equation}
\label{eq:VarExtrapolate_uniform}
    V_{L+1} = \left(M_{L+1}/M_L\right)^{-b_1}V_{L}. 
\end{equation}
%
\subsection{Adaptive multilevel Monte Carlo Finite-Element method}
\label{Sec:MLMC_AMR}
In view of the benefits of approximating the quantity of interest across a sequence of increasingly finer meshes, and with the goal of further reducing the computational cost associated with reducing the bias associated with the numerical discretization, it is natural to focus the refinement only on those parts of the mesh where the error is concentrated. Our goal is then to, starting from a computational mesh $\mathcal T_0$, generate a family of adaptively refined meshes $\left\{\mathcal T_\ell\right\}_{0\leq\ell\leq L}$ that will produce better approximations of $\mathbb E(u)$ than the ones resulting from consecutive uniform refinements of the initial grid. With this goal in mind, the use of an 
\textit{a posteriori} error estimator to guide the construction of the family of meshes has been proposed in the context of multilevel Monte Carlo methods \cite{EiMeNe:2016,HoVoSzTe:2012, HoVoSzTe:2014,KhPaHe:2020}.

\noindent \textbf{Residual error estimator.} A key ingredient in an adaptive solver is a local error estimator. In our case, for each element $K$ on the mesh $\mathcal{T}_\ell$, we will use the simple residual-based a posteriori error indicator
\begin{equation}
\label{eq:APosterioriErrEstimator}
        \eta_{K,\ell}(\boldsymbol\omega) := h_K^2 \left\Vert\, \nabla\cdot\left(\tfrac{1}{\mu r}\nabla u_\ell(\boldsymbol\omega)\right) - f(u_\ell(\boldsymbol\omega))\,\right\Vert_K + h_K^{3/2}\left\Vert\, \left[\!\left[\tfrac{1}{\mu r}\nabla u_\ell(\boldsymbol\omega)\cdot\boldsymbol n\right]\!\right]\,\right\Vert_{\partial K\setminus\partial\Omega},
\end{equation}
where $\partial K$ is the boundary of the element $K$, $h_K$ is the diameter of $K$, $\boldsymbol n$ is the outward unit normal to the element $K$, $[\![\cdot]\!]$ denotes the jump across the edge of an interior element, and $f$ is the source term defined piecewise on the right-hand side of \eqref{eq:FreeBoundarya}. Following \cite{EiMeNe:2016,KhPaHe:2020}, we will further define the mean local and mean global error estimators respectively as
\begin{equation}\label{eq:MeanEstimator}
\eta_{K,\ell}  := \mathbb E\left(\eta_{K,\ell}(\boldsymbol\omega)\right) \quad \text{ and } \quad \eta_\ell^2 := \sum_{K\in \mathcal{T}_\ell}\eta_{K,\ell}^2. 
\end{equation}
For linear deterministic problems, estimators of this form can be shown to be such that there are constants $C_1, C_2>0$ such that $C_1\eta_{\ell} \leq \|u - u_\ell\|_Z \leq C_2\eta_{\ell}$. Therefore, the error estimator will accurately locate the regions of high error density and will decay at the same rate as the true error \cite{LaTh:2003}. The global error can then be approximated by adding the local estimators over the entire triangulation. Using these error estimators, the adaptive analogue of \eqref{eq:EstimateV_l} can be written as
\begin{equation}\label{eq:AdaptiveEstimateV_l}
\mathbb{V}(u-u_\ell) \le \mathbb{E}\left[\left\Vert u-u_\ell\right\Vert_Z^2\right] \approx \eta_\ell^2,
\end{equation}
which then leads to the following adaptive analogue of the extrapolation formula \eqref{eq:VarExtrapolate_uniform}
\begin{equation}
\label{eq:AdaptiveVarExtrapolate}
    V_{L+1} = \left(\eta_{L+1}/\eta_L\right)^2V_{L}. 
\end{equation}
This estimate can then be used in combination with \eqref{eq:MLMC_SampleSize} to obtain an update for the sample size required at each adaptive level.\\ 

\textbf{Adaptive solution cycle.} With these definitions in place, we can then describe our strategy, which follows the ``SOLVE $\rightarrow$ ESTIMATE $\rightarrow$ MARK $\rightarrow$ REFINE" paradigm familiar from  deterministic adaptive solvers \cite{CaFePaPr:2014}, as:
\begin{enumerate}
\item \textbf{Solve:} Starting from a fixed number of samples, the problem \eqref{eq:FreeBoundarya} is solved on the initial mesh $\mathcal T_0$. 
\item \textbf{Estimate:} The local mean error estimator is approximated from the sample gathered. 
\item \textbf{Mark:} The set $\mathcal M_\ell$ containing the smallest possible number of elements in $\mathcal T_0$ satisfying
\begin{equation}
\label{eq: LocalRefineIndicator}
    \sum_{K\in \mathcal{M}_\ell}\eta_{K,\ell}^2\ge \zeta \eta_\ell^2,
\end{equation}
for some predetermined value $\zeta\in [0,1]$ is marked for refinement---this marking strategy is known as \textit{ D\"{o}rfler marking} in the adaptive finite element community \cite{Do:1996}.
\item \textbf{Refine:} The elements marked are refined in such a way that the resulting triangulation $\mathcal{T}_{\ell+1}$ is shape-regular and conforming. Efforts should be made to make sure that the growth of the number of elements is kept at bay. In our case, we used the implementation given in \cite{FuPrWi:2011} of the algorithms described in \cite{Carstensen:2003,FuPrWiTR:2008,MoNoKu:2000}.
\end{enumerate}
The steps above are repeated until the error estimator $\eta$ falls below a certain predetermined value. \\

\noindent \textbf{A notion of mesh level.} For uniformly refined grids, the notion of mesh level is natural: starting from a mesh $\mathcal T_\ell$, one step of uniform refinement decreases the mesh parameter $h$ across the grid by a factor of $1/2$; the resulting mesh is said to have level $\ell+1$ and is denoted by $\mathcal T_{\ell+1}$. For adaptively refined grids, where the mesh parameter is not constant through the grid, the notion of the level does not come so naturally. We will use the fact that, for a uniform refinement, the numerical error decays by a factor of $(1/2)^p$, (where $p$ is the order of the FEM solver) with each successive level to extend the notion of mesh level to adaptively refined grids.

Consider a numerical approximation $u_\ell$ obtained on a mesh $\mathcal T_\ell$ with an associated error estimation given by $\eta_\ell$. We will say that a mesh has level $\ell+1$ and will denote it by $\mathcal T_{\ell+1}$ if it was obtained from $\mathcal T_\ell$ by cycling over the adaptive loop using the value $(1/2)^p\eta_\ell$ as the stopping tolerance. In other words, we will say that an adaptively refined mesh has level $\ell+1$ if it produces a numerical solution with an error $(1/2)^p$ times smaller than one with level $\ell$, just like in the uniform case. We will refer to $q:=(1/2)^p$ as the decay factor. In terms of discretization accuracy, after $\ell$ steps of adaptive refinement, an adaptively refined mesh with level $\ell$ will have an associated error estimation $\eta_\ell=q^\ell\eta_0$, where $\eta_0$ corresponds to the error estimation at the initial mesh. In our numerical experiments, since the convergence rate of the piecewise linear solver is 2 (when measured in the $L^2$ norm), we shall use a decay factor $q=1/4$ to define our adaptively refined meshes.\\

\noindent \textbf{Deterministic adaptive grids}. Ideally, in the stochastic setting, all the error estimations collected from the totality of samples would be used to drive the adaptive refinement forward and build an optimal set of meshes at every level. However, due to the iterative nature of the algorithm arising from \eqref{eq:MLMC_SampleSize}, the optimal mesh at every level would have to be corrected with every new batch of samples and the solutions corresponding to all previous realizations $\boldsymbol\omega^{(i)}$ would have to be recomputed. The computational cost of re-sampling in this manner quickly becomes impractical. 

Instead, following \cite{HoVoSzTe:2012, HoVoSzTe:2014, KoYo:2018}, we will construct a sequence of deterministic adaptive grids with partial knowledge about $\mathbb E(u)$ as follows. Starting from a sample $\{\boldsymbol\omega^{(i)}\}_{1\leq i\leq N}$ (where $N$ is small and arbitrarily chosen) and a mesh $\mathcal T_0$, the PDE is solved and the local error is estimated for every solution $u_0^{(i)}$, resulting in $N$ local error estimators $\{\eta_{K,0}(\boldsymbol\omega^{(i)})\}_{1\leq i\leq N}$. The mean local and global error estimators $\eta_{K,0}$ and $\eta_0$ are then approximated by the sample means of the individual estimators. Using this approximation of $\eta_{K,0}$, the mesh $\mathcal T_0$ is refined. This process is continued until the approximated mean error estimator satisfies $\eta\leq q\eta_0$; the resulting mesh is stored and labeled as $\mathcal T_1$ (mesh level 1). The previous steps are repeated until a target number of meshes $\{\mathcal T_\ell\}_{0\leq\ell\leq L}$ have been generated. The process is described in Algorithm \ref{algo:MLMC_gen_adaptive_mesh}. Since the family of meshes produced is constructed using random samples of $\boldsymbol\omega$, they approximately reduce the error for the approximate expectation $\mathbb E(u_h)$ by a factor of $q$ with every increasing level. This family of meshes is then kept fixed during the MLMC run.

\begin{algorithm}[!ht]
\label{algo:MLMC_gen_adaptive_mesh}
\DontPrintSemicolon

    \KwIn{Initial mesh $\mathcal{T}_0$, maximum mesh level $L$, element marking percentage $\zeta\in (0,1)$, sample size $N$, error decay factor $q\in (0,1)$.}
    \KwOut{Adaptive mesh set $\{\mathcal{T}_\ell\}_{\ell = 0}^L$.}

        \For{$\ell= 1,\cdots,L$}
    {Refine = TRUE.
    
    Set $j=1$ and $\mathcal T^{(j)} = \mathcal T_{\ell -1}$.
    
    \While{Refine}{
    
    \For{$i = 1,\cdots,N $}{
    (i) Draw a random sample $\boldsymbol\omega^{(i)}$ from $W$.
    
    (ii) Obtain $u^{(i)}$ by solving the free boundary problem on on $\mathcal{T}^{(j)}$ using  $\boldsymbol\omega^{(i)}$.
    
    (iii) Calculate and accumulate the local and global error estimators $\eta_{K,j}(\boldsymbol{\omega}^{(i)})$ and $\eta_{j}(\boldsymbol{\omega}^{(i)})$.

    }
    Compute the mean estimators $\eta_{K,j}$ and $\eta_j$ from the accumulated samples.
    
    Mark \& refine the mesh $\mathcal{T}^{(j)}$ according to $\eta_{K,j}$ to obtain $\mathcal{T}^{(j+1)}$.
    
     \If{ $\eta_j\le q\eta_1$}
     {Store $\mathcal{T}^{(j)}$ as $\mathcal{T}_\ell$.
     
     Refine = FALSE.}

        $j=j+1$
    }
    }
\caption{Generate adaptive deterministic mesh set}
\end{algorithm}

\begin{algorithm}[!ht]
\DontPrintSemicolon

    \KwIn{Initial mesh level $L = 0,$ sequence of mesh available $\{\mathcal{T}_\ell\}_{\ell \ge 0}$, root nMSE $\epsilon$, $\zeta\in (0,1)$, initial sample size $N_{\text{old}}=\left\{N_\ell\right\}_{\ell = 0}^L$, counter $j=1$, sample size corrections $dN_\ell^j = \{N_\ell\}_{\ell=0}^L$.}
    \KwOut{$\left\{N_\ell\right\}_{\ell = 0}^L$, $\{V_\ell\}_{\ell = 0}^L$, $A_{\text{MLMC}}$.}
     \While{$\sum_\ell(dN_\ell^j)>0$}{
    \For{$0\le \ell\le L$}{
    
        \For{$i = 1,\cdots,dN_\ell^j $}
    {
    Solve the free boundary problem on $\mathcal{T}_\ell$ to get $u_\ell^{(i)}$ for the $i$-th sample.
    }
    }
     Approximate $\{V_\ell\}_{\ell=0}^L$ by \eqref{eq:MeanVarUpdate_Var}.
        
     Update the sample size estimation $\left\{N_\ell\right\}_{\ell = 0}^L$ by \eqref{eq:MLMC_SampleSize}. 
     
    $j=j+1$.
    
    $dN_\ell^j = \left\{N_\ell\right\}_{\ell = 0}^L-N_{\text{old}}.$
    
    $N_{\text{old}}=\left\{N_\ell\right\}_{\ell = 0}^L$.
    
    \If{$\sum_\ell(dN_\ell^j)=0$}{
    
    \If{ The a posteriori error estimator falls below $\sqrt{1-\theta}\epsilon \left\Vert\mathbb{E}(u) \right\Vert_{Z}$, }
    {
    Compute $A_{\text{MLMC}}$ by \eqref{eq:MLMC_estimator} and terminate the loop.
    }
    \Else {
     $L = L+1$.
     
     Approximate $V_L$ by \eqref{eq:VarExtrapolate_uniform} and compute $\left\{N_\ell\right\}_{\ell = 0}^L$ and go to Step 1.}
    }
    }  
\caption{Multilevel Monte Carlo Finite-Element method}\label{algo:MLMC_Algo_CorrectionVersion}
\end{algorithm}

\section{Numerical experiments}\label{sec:Num-Exp}
%
We now demonstrate and evaluate the performance of the methods presented for the Grad-Shafranov free boundary problem. We examine the efficiency of the three simulation approaches, comparing the CPU times (or computational cost) and the accuracy of some geometric descriptors that are generated from the approximation of $\mathbb E(\psi)$ obtained with each of the techniques.
Following \cite{FaHe:2017}, we consider an ITER geometry with 12 coils and a ``baseline'' vector of target current intensities $\boldsymbol{I}$ given by
\begin{equation}\label{eq:CentralCurrentValue}
{\renewcommand{\arraycolsep}{2pt}
\begin{array}{llll}
I_1 = -1.4 \times 10^{6} A, \quad & I_2 = -9.5 \times 10^{6} A, \quad & I_3 = -2.0388 \times 10^{7} A, \quad & I_4 = -2.0388 \times 10^{7} A, \\
I_5 = -9 \times 10^{6} A, \quad & I_6 = 3.564 \times 10^{6} A, \quad & I_7 = 5.469 \times 10^{6} A, \quad & I_8 = -2.266 \times 10^{6} A, \\
I_9 = -6.426 \times 10^{6} A, \quad & I_{10} = -4.82 \times 10^{6} A, \quad & I_{11} = -7.504 \times 10^{6} A, \quad & I_{12} = 1.724 \times 10^{7} A. 
\end{array}
}
\end{equation}
We will refer to these values as the \textit{reference currents}. The  profiles for $p$ and $g$ on the right hand side of \eqref{eq:FreeBoundarya} follow the form given in \eqref{eq:source} with the specific values $r_0=6.2m,\,\beta=0.5978, \, \alpha_1 = 2$, and $\alpha_2=1.395$, and $j_0=1.3655 \times 10^6 A/m^2$. The reactor and coil array geometries follow the ones described in \cite{Amoskov2009}. In our experiments, we will take the vector of current intensities to be subject to uncertainty modeled as a uniformly distributed perturbation of magnitude $\tau=2\%$ centered around the reference values above.

\subsection{Experiment description}
In this section, we present numerical results comparing the three approaches -- MC-FE, uniform MLMC-FE, and adaptive MLMC-FE. For the solution of \eqref{eq:FreeBoundary}, we used the finite element-based solver {\tt FEEQS.m} \cite{Heumann:feeqsm} developed by Holger Heumann and collaborators as a lightweight Matlab implementation of the code {\tt CEDRES++} \cite{FaHe:2017,CEDRES}. The code implements a piecewise linear finite element discretization of a weak formulation of \eqref{eq:FreeBoundary} and employs a globalized variation of Newton's method to resolve the nonlinearity.  (The stopping threshold for the relative residual was set to $5\times 10^{-11}$.) For the solution of the perturbed problems, the initial iterate of Newton's method was taken to be the solution corresponding to the reference currents $\boldsymbol{I}$. All tests used the splitting parameter $\theta=0.5$ in \eqref{eq:nMSE_MC_ErrSplitting}. The user-specified tolerances for the normalized mean squared error range from $\epsilon=2\times 10^{-4}$ to $8\times 10^{-3}$.
Experiments were conducted using {\tt MATLAB} R2022a on a System 76 Thelio Major with 256GB RAM and a 64-Core @4.6 GHz AMD Threadripper 3 Processor.

\begin{figure}[ht!]\centering
\begin{tabular}{cccc}
\includegraphics[width=0.45\linewidth]{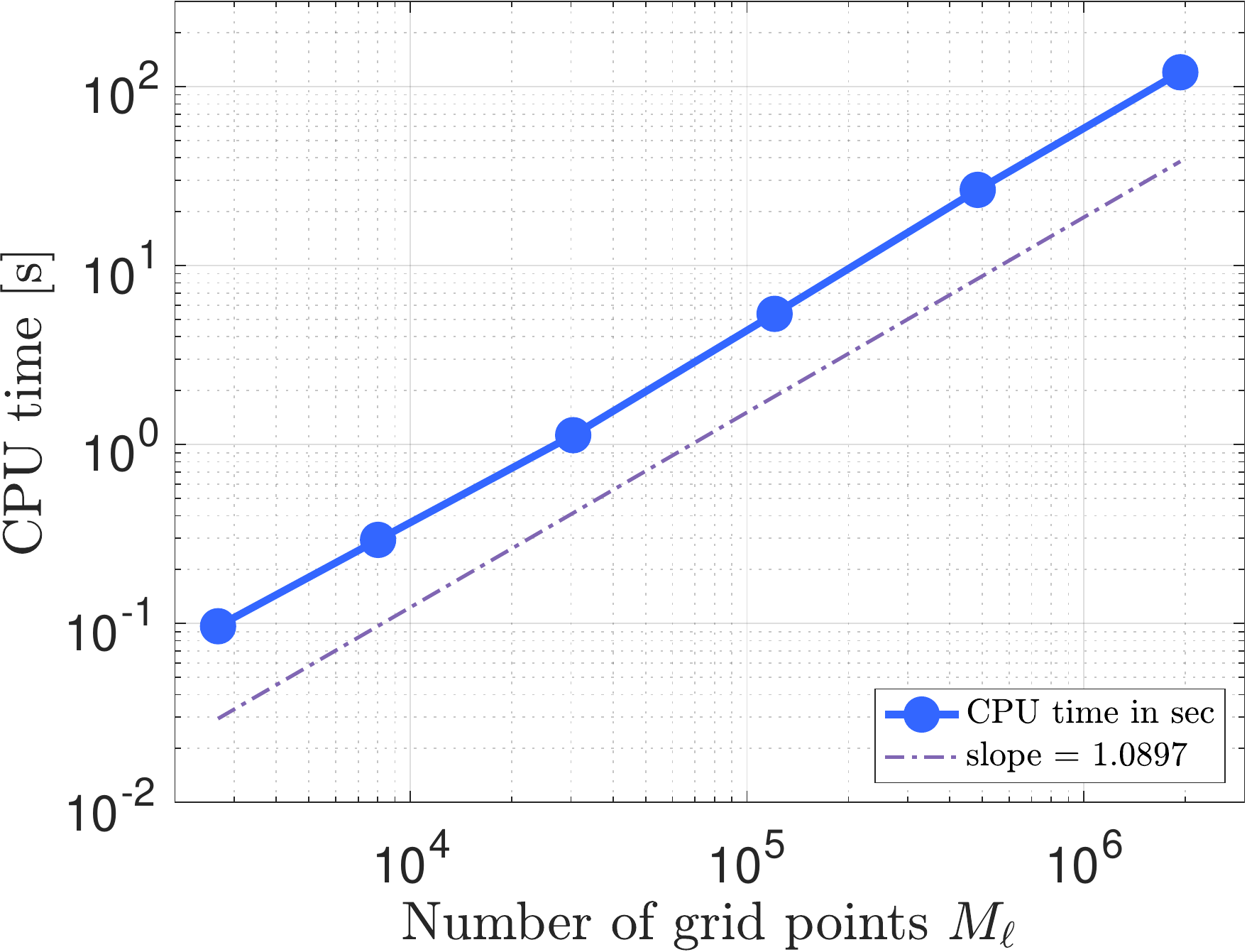}
\end{tabular}
\caption{Mean CPU time to compute 100 realizations of $u_\ell$, as a function of the number of grid points $M_\ell$, plotted 
on a logarithmic scale.  The fitted curve indicates that the computational cost $C_\ell$ behaves approximately like $M_\ell^{1.09}$.}
\label{fig:CostEstimatePlot} 
\end{figure}

To produce an estimate of the number of samples required on each discretization level, equation \eqref{eq:MLMC_SampleSize} requires the knowledge of two problem-dependent parameters: the power $c$ appearing in the estimate for the computational cost \eqref{eq:AverageCost}, and the normalized variance of the correction terms $V_\ell$ as defined in \eqref{eq:NormalizedVariance}. The normalization factor $\|\mathbb{E}(u_L)\|_Z$ in \eqref{eq:NormalizedVariance} was estimated on the finest uniform mesh level ($\ell=5$) to be approximately $8.57\times 10^{-1}$. To estimate the value of $c$, 100 random currents are sampled for different mesh sizes $M_\ell$, the processing times required to obtain the solutions are averaged for each mesh size, and $c$ is estimated through a regression. Figure \ref{fig:CostEstimatePlot} shows the behavior of the average cost as a function of the mesh size $M_\ell$; from the data displayed, the power law is estimated to be $c \approx 1.09$. Note that this cost estimate is based on Matlab timings and not on the complexity analysis of standard linear solution algorithms. The same samples are also used to estimate the sample means, $\mathbb E(Y_\ell)$ or $\mathbb E(u_h)$, and variances, $\mathbb V(Y_\ell)$ or $\mathbb V(u_h)$ dynamically using  Welford's algorithm \cite{Welford:1962}. As the new samples are gathered, the mean $m_w$ and proxy for the variance, $s_w$, are updated using the following formulas for the $i$-th sample, with $m_w^{(0)}=0, s_w^{(0)}=0$:
\[
m_w^{(i)}=m_w^{(i-1)} + \frac{u^{(i)}-m_w^{(i-1)}}{i},\quad s_w^{(i)} = s_w^{(i-1)} + \left\langle u^{(i)}-m_w^{(i-1)},u^{(i)}-m_w^{(i)} \right\rangle .
\]
Using sample size $i$, the variance is then given by $V^{(i)} = s_w^{(i)}/(i-1)$.

\begin{table}[ht]
	\centering
	\scalebox{0.8}{
		\begin{tabular}{c|c|c|c|c|c|c|c|c|c|c|c|c|c|c|c|c|c|c|}
			\cline{1-7}	
			\multicolumn{1}{|c|}{ Level $\ell$} &0&1&2&3&4&5\\
			\hline
   \multicolumn{1}{|c|}{Uniform}&2685&8019&30449&120697&484080&1934365\\
   \hline

  \multicolumn{1}{|c|}{Adaptive}&2685  &6090       &25099      &103968      &411913 &1552282\\
      \hline
	\end{tabular}}
			\caption{The number of grid points $M_\ell$ for both geometry-conforming uniform and adaptive ($q=1/4$) meshes as the mesh levels increase from 0 to 5.}
			\label{Tab:Dof}
\end{table}

To perform MLMC-FE simulations, the user typically defines and generates a sequence of spatial grids, where, given a tolerance $\epsilon$, the fineness of the grid is determined by the requirement that the discretization error ($\widehat{\mathcal{E}}_{\text{Bias}}^2$ in \eqref{eq:nMSE_MC_ErrSplitting}) or an estimate of it be small enough.
In this study, we generated two types of grids, a set of {\it geometry-conforming} uniformly refined grids, and a set of adaptively refined grids constructed using the strategy presented in Section \ref{Sec:MLMC_AMR}.\footnote{The domain components contain curved boundaries, which we handled by treating them as polygonal structures.  The mesh generation entails identifying the curved boundaries using piecewise splines and interpolating along these splines with grids of varying fineness.}  
For the uniformly refined grid, we generated a total of six levels of grids. We note that, due to the increasing accuracy of the spline approximation of the curved boundaries, these meshes are not a direct refinement of each other. Instead, each level is characterized by a mesh parameter $h_\ell$ being roughly half of the preceding mesh as determined by the {\tt Triangle} mesh generator \cite{Sh:2002}.
The adaptive refinement strategy began with the coarsest mesh from the uniform family and applied the weighted $L_2$ a posteriori error estimator specified in \eqref{eq:APosterioriErrEstimator} and $q=1/4$ to reflect a similar error decay as for uniform refinement, also using {\tt Triangle}  to generate the desired adaptively refined meshes.
The number of grid points for each of these methods on different grid levels is shown in Table \ref{Tab:Dof}. Note that the grid sizes for the adaptive meshes are not dramatically smaller than for uniform meshes, which suggests that the solution, as a function of the spatial variables, is fairly smooth.

%
\subsection{Computational cost}
%

\begin{figure}[ht!]\centering
\begin{tabular}{cc}
\includegraphics[width=0.48\linewidth]{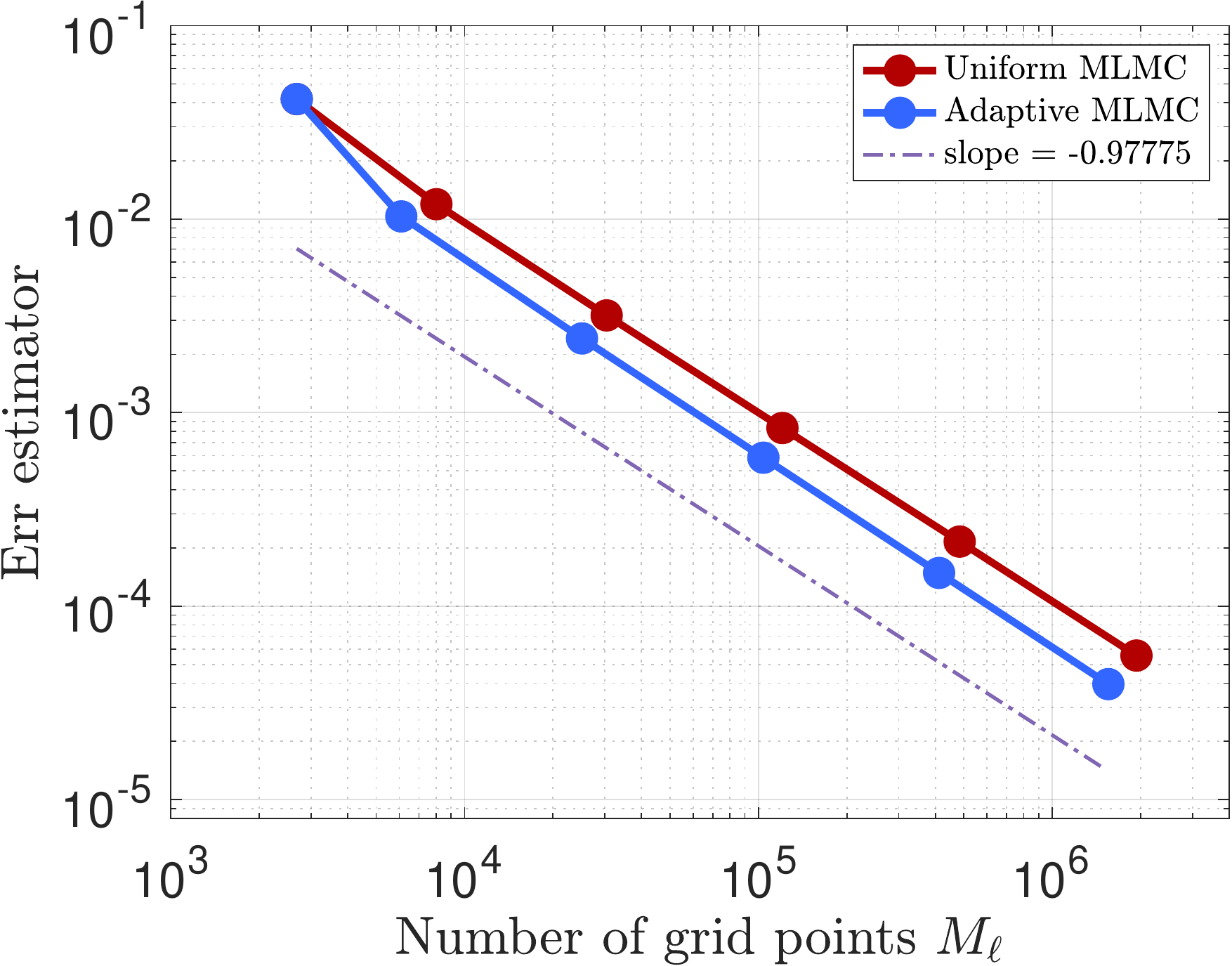}&
\includegraphics[width=0.48\linewidth]{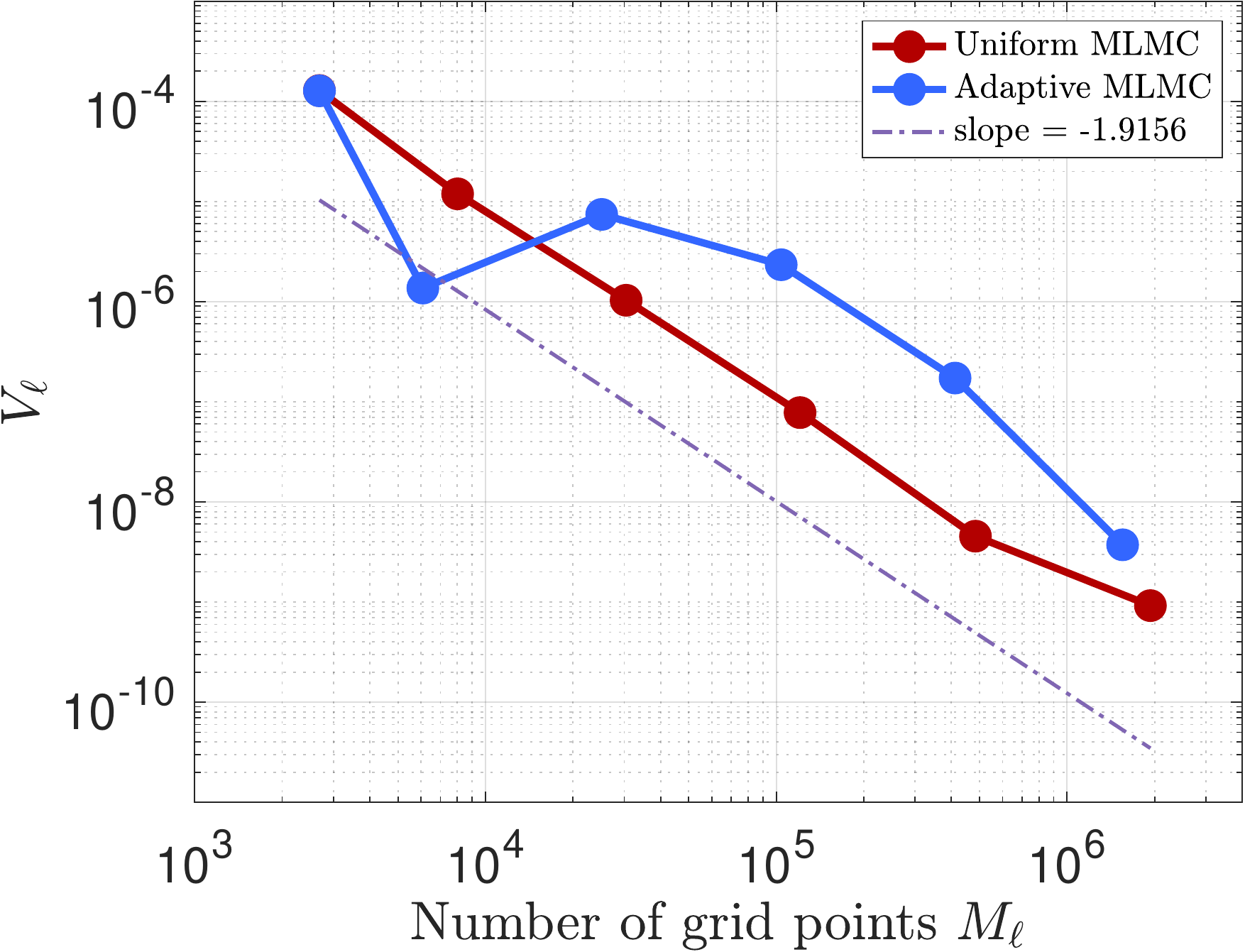}\\
\includegraphics[width=0.48\linewidth]{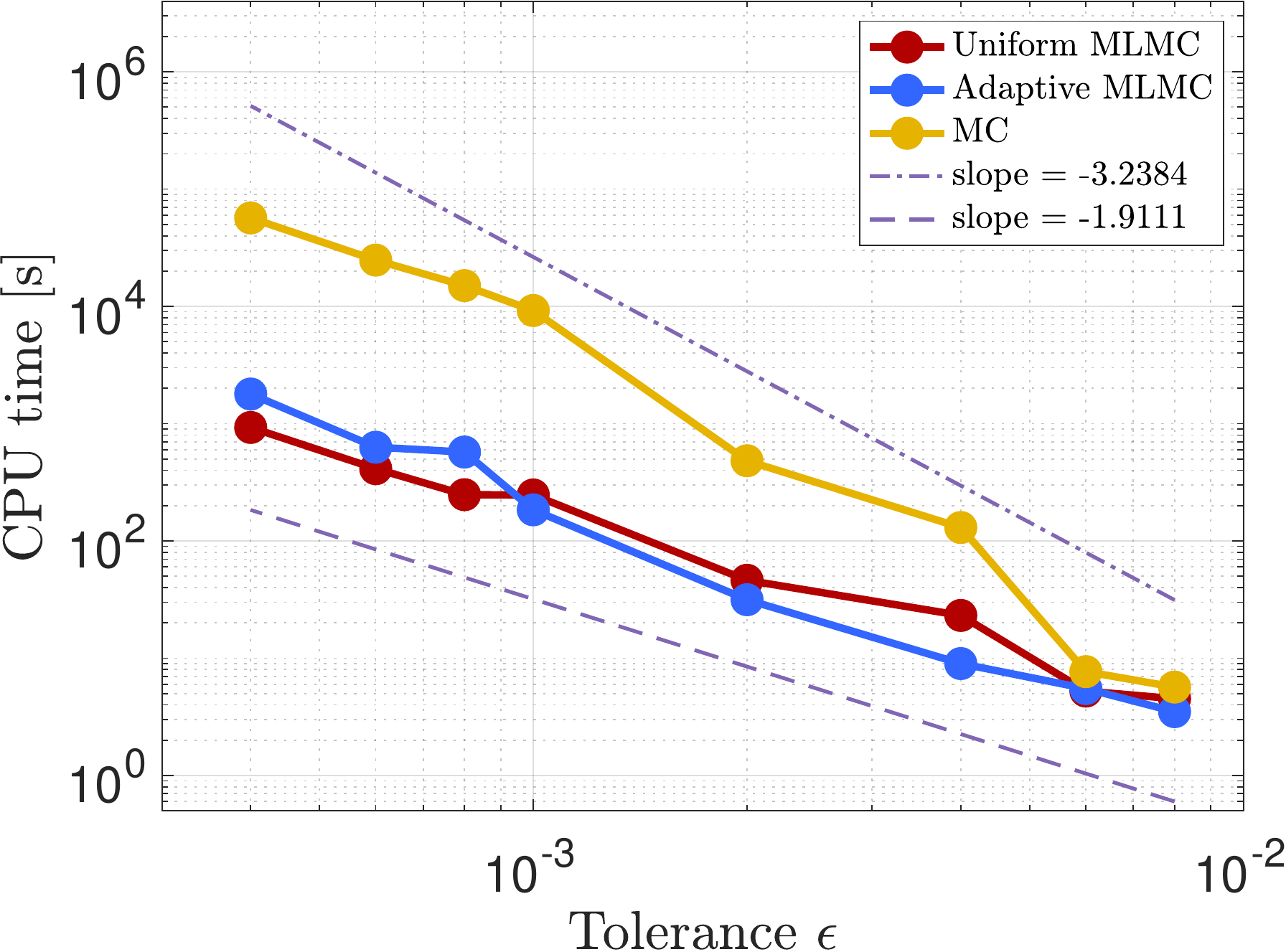}&
\includegraphics[width=0.48\linewidth]{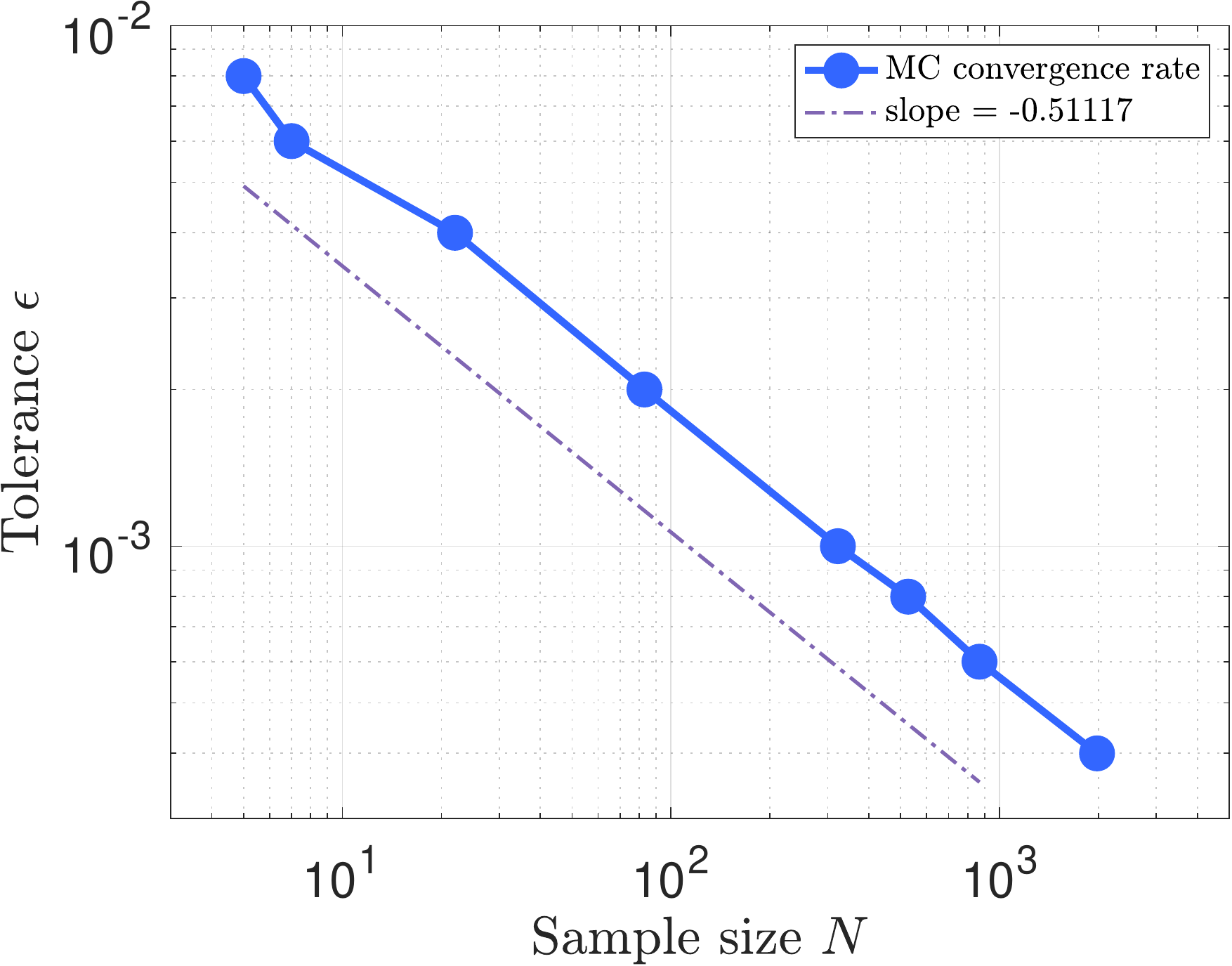}
\end{tabular}
\caption{Top left: weighted $L_2$ error (with weight $\mu x$) of estimator $C\eta_\ell$ vs. number of grid points $M_\ell$ plot.  Top right: normalized variance $V_\ell$ vs. number of grid points $M_\ell$ plot.  Bottom left: CPU time in seconds vs. tolerance $\epsilon$. Bottom right: Monte Carlo convergence rate estimate with tolerance $\epsilon$ vs. sample size $N$. This plot is generated from Table \ref{Tab:SampleSize}.} 
\label{fig:Experiment_result_plot}
\end{figure}

Figure \ref{fig:Experiment_result_plot} shows a variety of computational results, including the error estimator, $V_\ell$, and CPU time for the two versions of MLMC (and times for full MC). To investigate the convergence behavior of the discretization error, we calculate the a posteriori error estimator for both uniform and adaptive meshes in the same experiment to obtain an estimate of $c$ for $C_\ell$ before conducting the simulations. The results are displayed in the top left plot in Figure \ref{fig:Experiment_result_plot}, with a dashed line showing a least square fit, indicating that the discretization error of both methods exhibits an asymptotic rate of $\mathcal{O}(M^{-1})$ (or $p \approx 2$).  The similar convergence rate further indicates that the solution to the problem is smooth and the error is equidistributed, rendering the adaptive strategy comparable to uniform mesh refinement. Note that the estimated error is used for variance extrapolation in \eqref{eq:AdaptiveVarExtrapolate} during the MLMC simulations.

The top right plot of Figure \ref{fig:Experiment_result_plot} shows the behavior of $V_\ell$ of \eqref{eq:NormalizedVariance} for both uniform and adaptive MLMC-FE methods with $\epsilon=2\times 10^{-4}$, using six levels of meshes. It can be seen that both methods demonstrate a decreasing trend in the values of $V_\ell$ as the mesh resolution increases, with a power law decay characterized by $b \approx 2$ in the least square fit. But there is a regime for a small number of grid points where the asymptotic behavior of the adaptive method is not evident, in contrast to the behavior of the uniform method. As the meshes get finer, the plots of $V_\ell$ for the two methods are close to being parallel. 

The computational effort for uniform MLMC-FE and MC-FE scales as $\mathcal{O}(\epsilon^{-2})$ and $\mathcal{O}(\epsilon^{-3})$ respectively, as indicated by the slopes of the least square fitting lines for the red and yellow curves. This observation is consistent with the theoretical cost predictions in Theorem \ref{thm:Theorem2} (with $b>c$) and \eqref{eq:MC_Cost}.  Theorem \ref{thm:Theorem2} also indicates that the majority of computational work is performed on coarse grids. 
Table \ref{Tab:SampleSize} shows the sample sizes obtained from \eqref{eq:MC_SampleSizeEstimate} for MC-FE and \eqref{eq:MLMC_SampleSize} for
MLMC-FE, further demonstrating a decrease in $N_\ell C_\ell$ as $\ell$ increases for
the multilevel methods.
We also found that the computational cost associated with the smallest tolerance $\epsilon = 2\times 10^{-4}$ is so large that we were unable to generate MC-FE results on a fine mesh ($\ell=5$) with a large sample size. In contrast, both versions of MLMC-FE successfully generated results with this tolerance.\footnote{Although we could not directly generate results for MC-FE for $\epsilon=2\times 10^{-4}$, we could estimate the costs.
In particular, we found that the variance $V_h$ is close to constant across mesh levels.  
Consequently, we used \eqref{eq:MC_SampleSizeEstimate} to estimate the number of required samples as 8000 in Table \ref{Tab:SampleSize}, approximately four times the number required for $\epsilon=4\times  10^{-4}$. This number was multiplied by the mean CPU time observed for the computations for Figure \ref{fig:CostEstimatePlot} (120.3 seconds, the largest entry appearing in the figure) to 
give the estimated total CPU time in Table \ref{Tab:CPU_time}.
 Table  \ref{Tab:CPU_time} gives quantitative data on the costs in CPU time for the three methods, as well as the 
speedups achieved by the two multilevel methods. 
It can be seen that for small tolerance $\epsilon$, both these methods reduce the CPU times dramatically, with many examples of speedups greater than a factor of 60 and a best-case speedup of approximately 200.}

As seen in the bottom left plot of Figure \ref{fig:Experiment_result_plot}, the uniform and adaptive versions of MLMC-FE have similar computational costs of $\mathcal{O}(\epsilon^{-2})$, as evidenced by the similar decay rate of the error estimator and \eqref{eq:AdaptiveEstimateV_l}. According to \eqref{eq:MLMC_Cost}, the slightly smaller magnitude of the error estimator for the adaptive MLMC-FE suggests a smaller (or comparable) computational cost in the asymptotic regime. However, when $\epsilon=4\times 10^{-4}$, the adaptive MLMC-FE method requires approximately twice as much CPU time ($1.79\times 10^{3}$ seconds) compared to the uniform MLMC-FE approach ($9.29\times 10^{2}$ seconds) due to a notable increase in $V_\ell$ around $M_\ell=10^4$. This also causes the speedups achieved using adaptive refinement to be somewhat smaller than for uniform refinement.
Thus, the traditional advantage of adaptive mesh refinement is not clearly present. We also attribute this to the apparent smoothness of the solution. We will demonstrate some advantages of the adaptive strategy in Section \ref{sec:geoparms}.

The bottom right plot of Figure \ref{fig:Experiment_result_plot} shows that the nMSE tolerance $\epsilon$ of MC-FE approach declines at $\mathcal{O}(N^{-0.51})$, which is consistent with the well-known square root convergence rate. This rate holds since $\mathbb{V}(u_h)$ remains nearly constant among all levels. 

\begin{table}[ht]
	\centering
			\scalebox{0.62}{
   \begin{tabular}{c|c|c|c|c|c|c|c|c|c|c|c|c|}
	    \cline{2-7}	
		&\multicolumn{6}{|c|}{ Level $\ell$}\\
			\hline
			\multicolumn{1}{|c|}{$\epsilon$}&0&1&2&3&4&5\\
			\hline
			\multicolumn{1}{|c|}{$8\times 10^{-3} $}&&&5&&&\\
			\multicolumn{1}{|c|}{$6\times 10^{-3} $}&&&7&&&\\
			\multicolumn{1}{|c|}{$4\times 10^{-3} $}&&&&22&&\\
			\multicolumn{1}{|c|}{$2\times 10^{-3} $}&&&&83&&\\
			\multicolumn{1}{|c|}{$10^{-3} $}&&&&&322&\\
			\multicolumn{1}{|c|}{$8\times 10^{-4} $}&&&&&527&\\
			\multicolumn{1}{|c|}{$6\times 10^{-4} $}&&&&&869&\\
                \multicolumn{1}{|c|}{$4\times 10^{-4} $}&&&&&1980&\\
                \multicolumn{1}{|c|}{$2\times 10^{-4} $}&&&&&& 8000$^{\ast}$\!\!\\
			\hline
	\end{tabular}
 \qquad
		\begin{tabular}{c|c|c|c|c|c|c|c|c|c|c|c|c|}
	    \cline{2-7}	
		&\multicolumn{6}{|c|}{ Level $\ell$}\\
			\hline
			\multicolumn{1}{|c|}{$\epsilon$}&0&1&2&3&4&5\\
			\hline
			\multicolumn{1}{|c|}{$8\times 10^{-3} $}&10     &2     &2&&&\\
			\multicolumn{1}{|c|}{$6\times 10^{-3} $}&12     &3     &2&&&\\
			\multicolumn{1}{|c|}{$4\times 10^{-3} $}&32     &5     &2     &2&&\\
			\multicolumn{1}{|c|}{$2\times 10^{-3} $}&152    &26     &4     &2&&\\
			\multicolumn{1}{|c|}{$10^{-3} $}&691   &109    &18     &4     &2&\\
			\multicolumn{1}{|c|}{$8\times 10^{-4} $}&841   &129    &23     &3     &2&\\
			\multicolumn{1}{|c|}{$6\times 10^{-4} $}&1610         &231          &40           &8           &2&\\
                \multicolumn{1}{|c|}{$4\times 10^{-4} $}&3791         &589         &104          &15           &3&\\
                \multicolumn{1}{|c|}{$2\times 10^{-4} $}&15859        &2344         &375          &62          &13           &2\\
			\hline
	\end{tabular}
 \qquad
		\begin{tabular}{c|c|c|c|c|c|c|c|c|c|c|c|c|c|c|c|c|c|}
	    \cline{2-7}	
		&\multicolumn{6}{|c|}{ Level $\ell$}\\
			\hline
			\multicolumn{1}{|c|}{$\epsilon$}&0&1&2&3&4&5\\
			\hline
			\multicolumn{1}{|c|}{$8\times 10^{-3} $}&10     &2     &2&&&\\
			\multicolumn{1}{|c|}{$6\times 10^{-3} $}&19     &3     &3&&&\\
			\multicolumn{1}{|c|}{$4\times 10^{-3} $}&38     &5     &4&&&\\
			\multicolumn{1}{|c|}{$2\times 10^{-3} $}&121    &18     &6     &2&&\\
			\multicolumn{1}{|c|}{$10^{-3} $}&697    &82    &49     &8&&\\
			\multicolumn{1}{|c|}{$8\times 10^{-4} $}&1446         &118          &91          &27           &6&\\
   \multicolumn{1}{|c|}{$6\times 10^{-4} $}&2070         &218         &133          &21           &3&\\
			\multicolumn{1}{|c|}{$4\times 10^{-4} $}&5075         &484         &315          &61          &14&\\
                \multicolumn{1}{|c|}{$2\times 10^{-4} $}&25871        &1961        &1668         &430          &85          &14\\
			\hline
	\end{tabular}
 
 }
	\caption{The optimal sample size estimation for MC-FE (left), uniform MLMC-FE (middle), and adaptive MLMC-FE (right). The simulations were conducted for a variety of choices of $\epsilon$. The computational cost associated with a tolerance of $\epsilon = 2\times 10^{-4}$ for Monte Carlo was prohibitive; the entry in the table for this tolerance (with an asterisk) is an estimate.}
	\label{Tab:SampleSize}
\end{table}

\begin{table}[ht]
	\centering
			\scalebox{0.62}{
   \begin{tabular}{c|c|c|c|c|c|c|c|c|c|c|c|c|}
			\hline
			\multicolumn{1}{|c|}{ }&MC-FE& Uniform MLMC-FE&Adaptive MLMC-FE\\
			\multicolumn{1}{|c|}{$\epsilon$}&Time & \begin{tabular}{cc} \,\,\,\,\,Time & \,\,\,Speedup \end{tabular}  &\begin{tabular}{cc} \,\,\,\,Time & \,\,\,Speedup \end{tabular}\\
			\hline
			\multicolumn{1}{|c|}{$8\times 10^{-3} $}&5.67e+00&\begin{tabular}{cc}4.52e+00\,\,\, & 1.3 \end{tabular}&\begin{tabular}{cc}3.50e+00\,\,  & 1.6 \end{tabular} \\
			\multicolumn{1}{|c|}{$6\times 10^{-3} $}&7.69e+00&\begin{tabular}{cc}5.25e+00\,\,\, & 1.5 \end{tabular}&\begin{tabular}{cc}5.51e+00\,\, & 1.4 \end{tabular}\\
			\multicolumn{1}{|c|}{$4\times 10^{-3} $}&1.30e+02&\begin{tabular}{cc}2.32e+01\,\,\, & 5.6 \end{tabular}&\begin{tabular}{cc}9.01e+00& 14.4 \end{tabular}\\
			\multicolumn{1}{|c|}{$2\times 10^{-3} $}&4.83e+02&\begin{tabular}{cc}4.62e+01\,\, & 10.5 \end{tabular}&\begin{tabular}{cc}3.16e+01& 15.3 \end{tabular}\\
			\multicolumn{1}{|c|}{$10^{-3} $}&9.22e+03&\begin{tabular}{cr}2.47e+02\,\, & 37.3 \end{tabular}&\begin{tabular}{cc}1.84e+02& 50.1 \end{tabular}\\
			\multicolumn{1}{|c|}{$8\times 10^{-4} $}&1.50e+04&\begin{tabular}{cc}2.48e+02\,\, & 60.5 \end{tabular}&\begin{tabular}{cc}5.73e+02& 26.2 \end{tabular}\\
			\multicolumn{1}{|c|}{$6\times 10^{-4} $}&2.48e+04&\begin{tabular}{cc}4.13e+02\,\, & 60.0 \end{tabular}&\begin{tabular}{cc}6.30e+02& 39.4 \end{tabular}\\
                \multicolumn{1}{|c|}{$4\times 10^{-4} $}&5.68e+04&\begin{tabular}{cc}9.29e+02\,\, & 61.1 \end{tabular} &\begin{tabular}{cc}1.79e+03& 31.7 \end{tabular}\\
                \multicolumn{1}{|c|}{$2\times 10^{-4} $}&9.62e+05$^{\ast}$\!\!\!&\begin{tabular}{cc} 4.21e+03 & 228.5 \end{tabular} &\begin{tabular}{cc}1.21e+04& 79.6 \end{tabular}\\
			\hline
	\end{tabular}
 }
	\caption{The CPU time in seconds for MC-FE (left), uniform MLMC-FE (middle), and adaptive MLMC-FE (right), together with speedups for the multilevel methods, for a variety of choices of $\epsilon$. The computational cost associated with a tolerance of $\epsilon = 2\times 10^{-4}$ for Monte Carlo was prohibitive; the entry in the table for this tolerance (with an asterisk) is an estimate.}
	\label{Tab:CPU_time}
\end{table}

%
\subsection{Properties of geometric parameters}  \label{sec:geoparms}
%
Next, we will explore the plasma boundaries and geometric descriptors of the expected poloidal flux $\psi$ resulting from the three methods. To ensure a fair comparison, we will use the results obtained from the MC-FE on the finest uniform mesh as a reference benchmark.\\

\begin{figure}[ht!]\centering
\begin{tabular}{ccccc}
\includegraphics[width=0.19\linewidth]{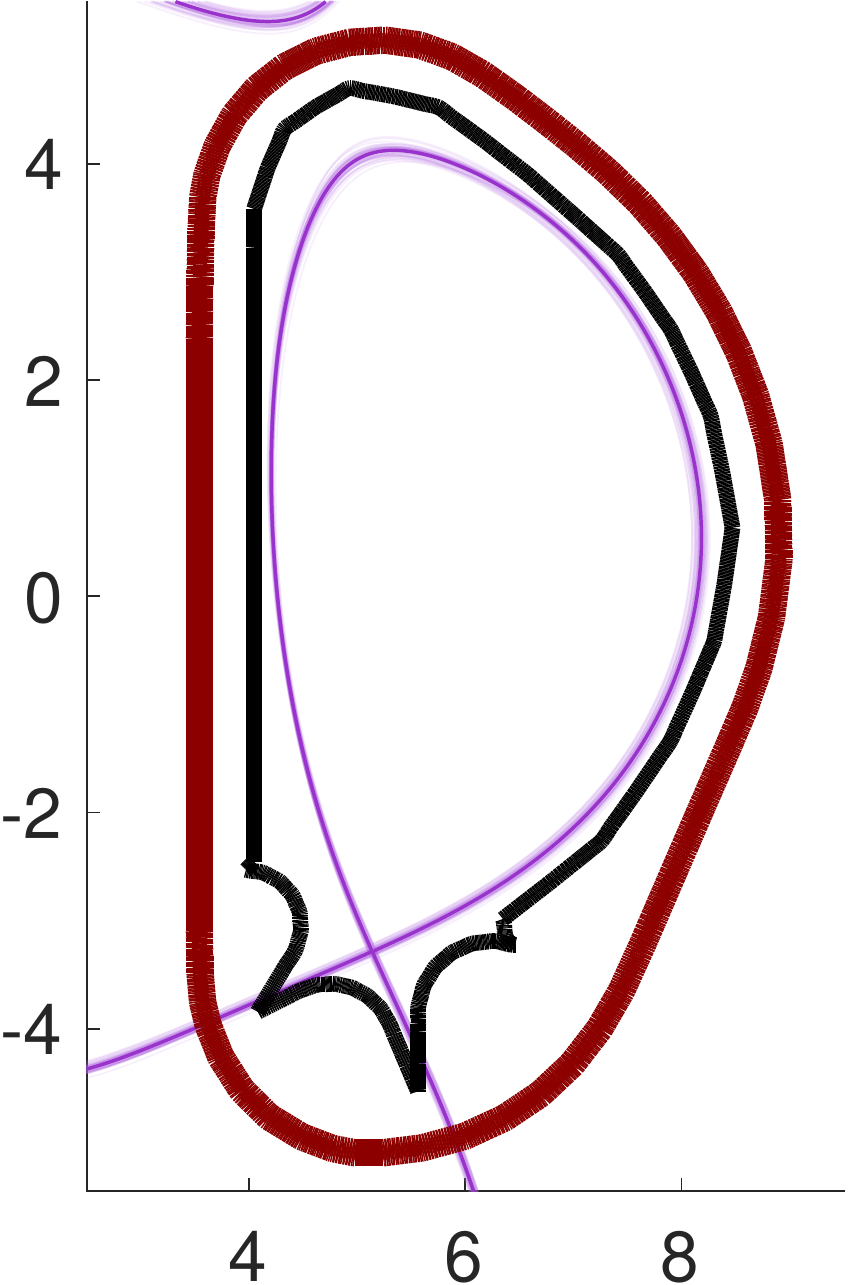} &
\qquad &
\includegraphics[width=0.19\linewidth]{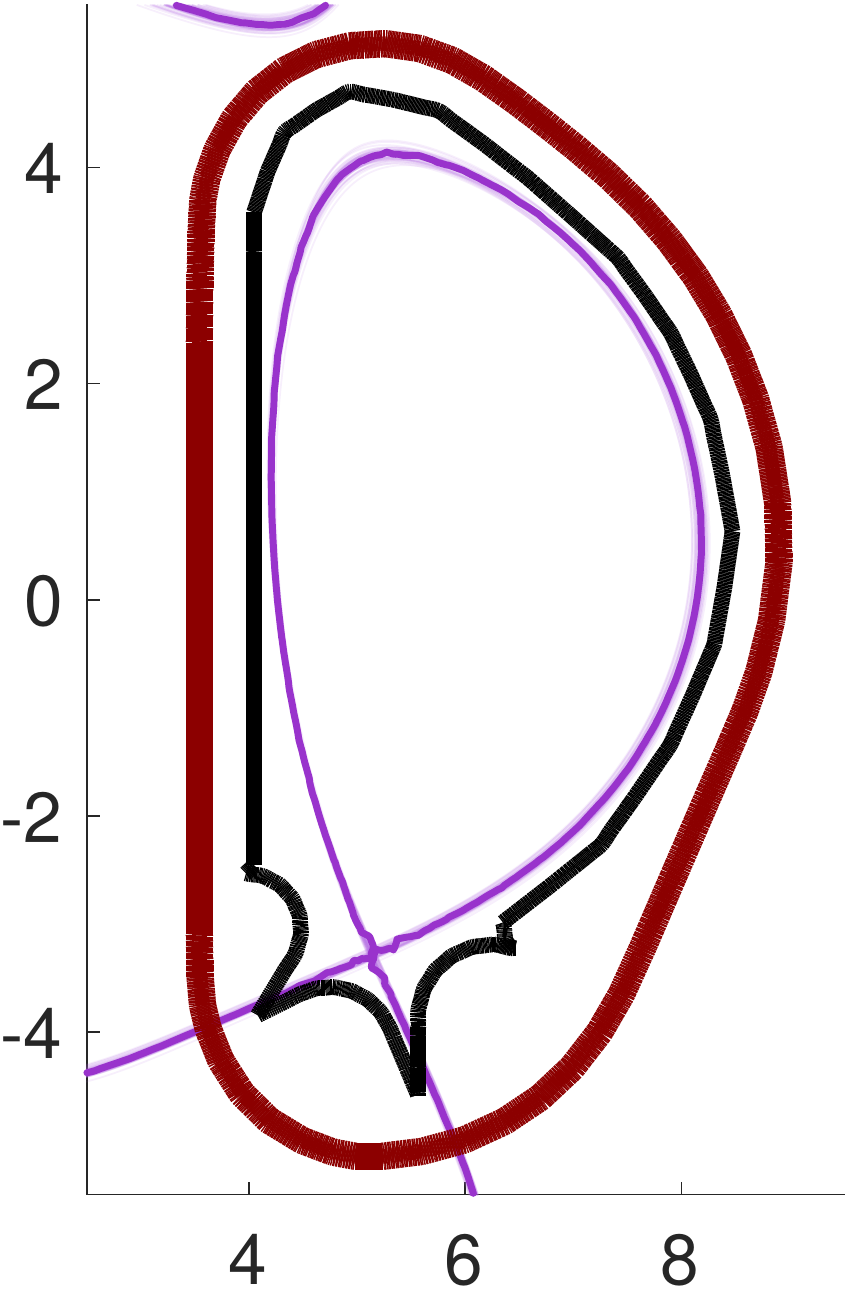} &
\qquad &
\includegraphics[width=0.19\linewidth]{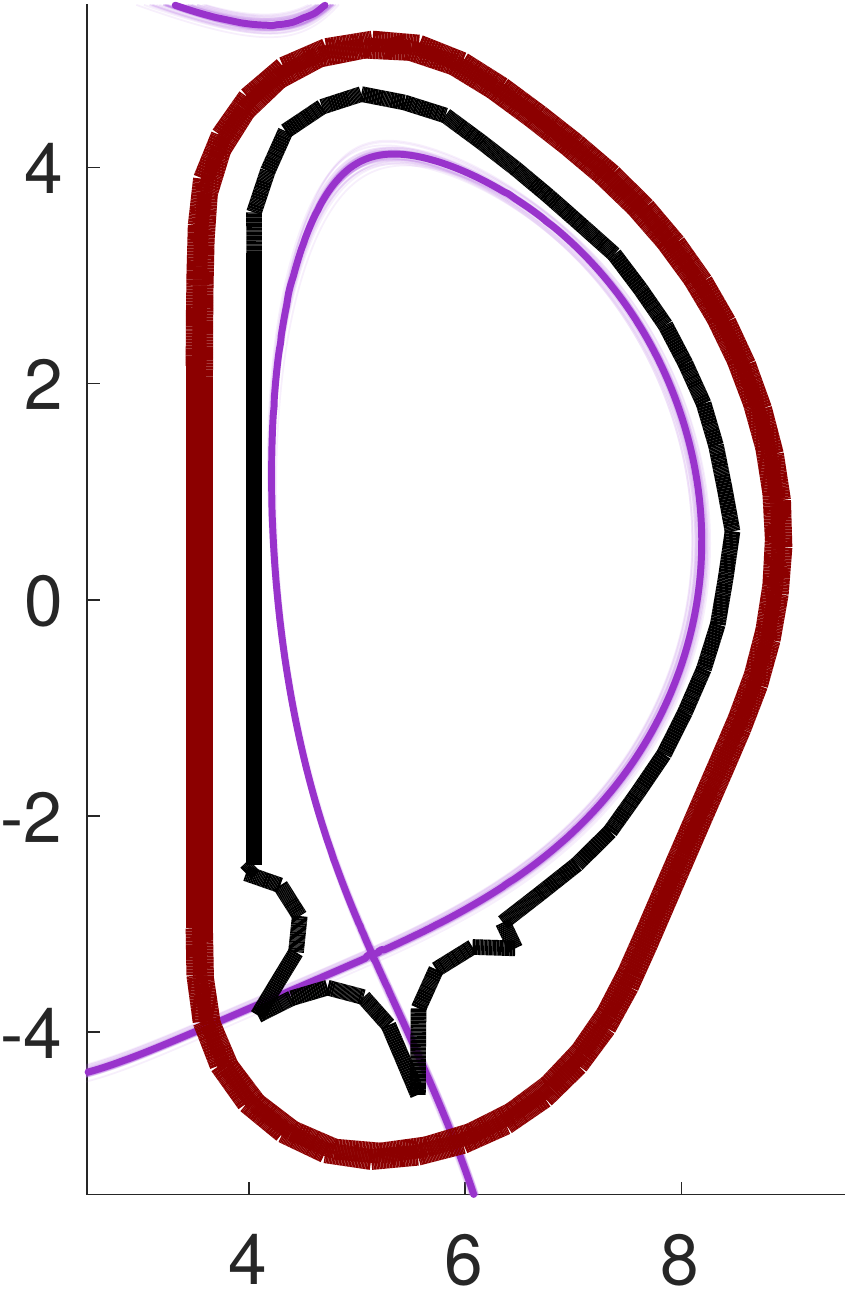} \\
\includegraphics[width=0.19\linewidth]{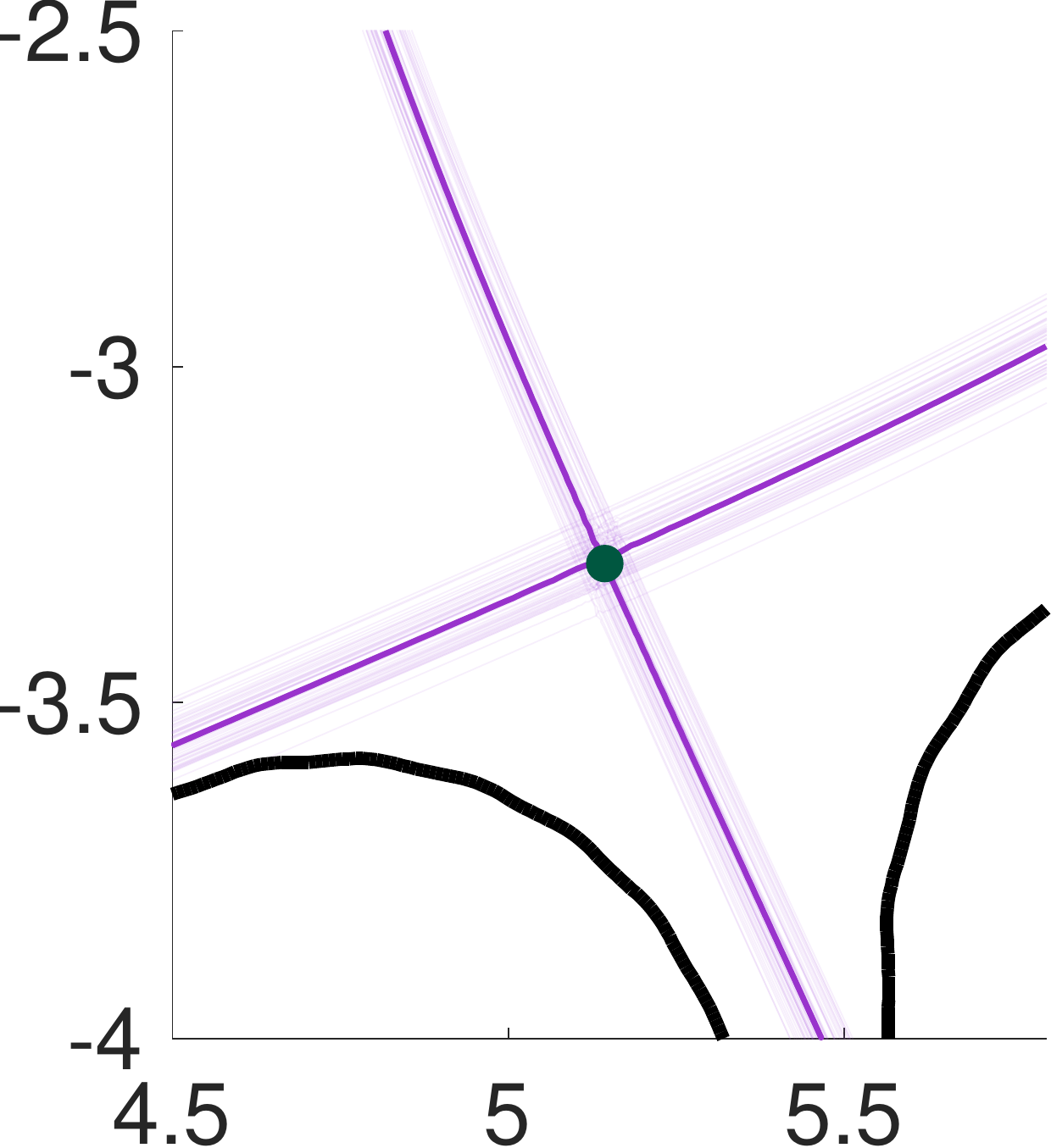} &
\qquad &
\includegraphics[width=0.19\linewidth]{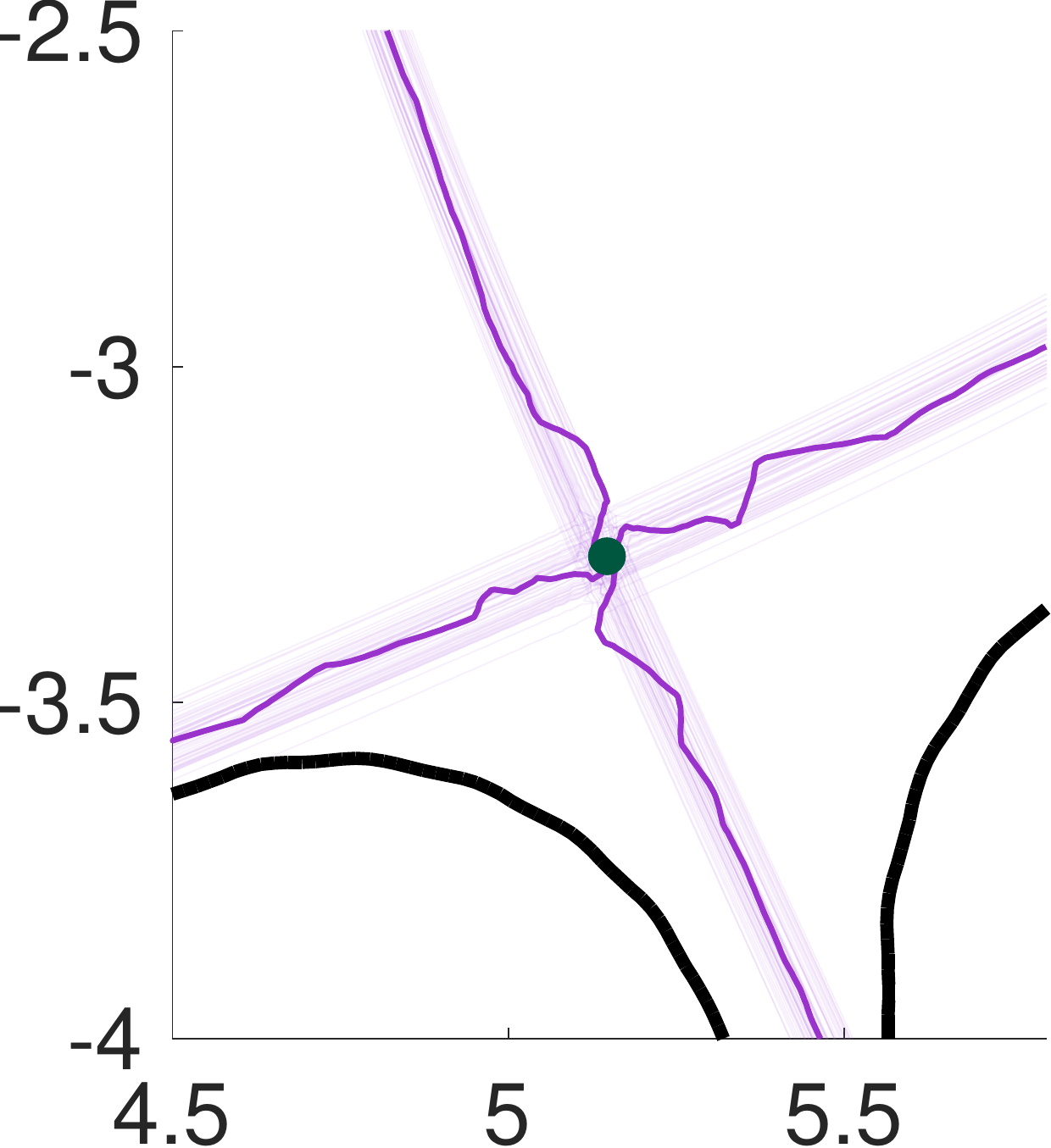} &
\qquad &
 \includegraphics[width=0.19\linewidth]{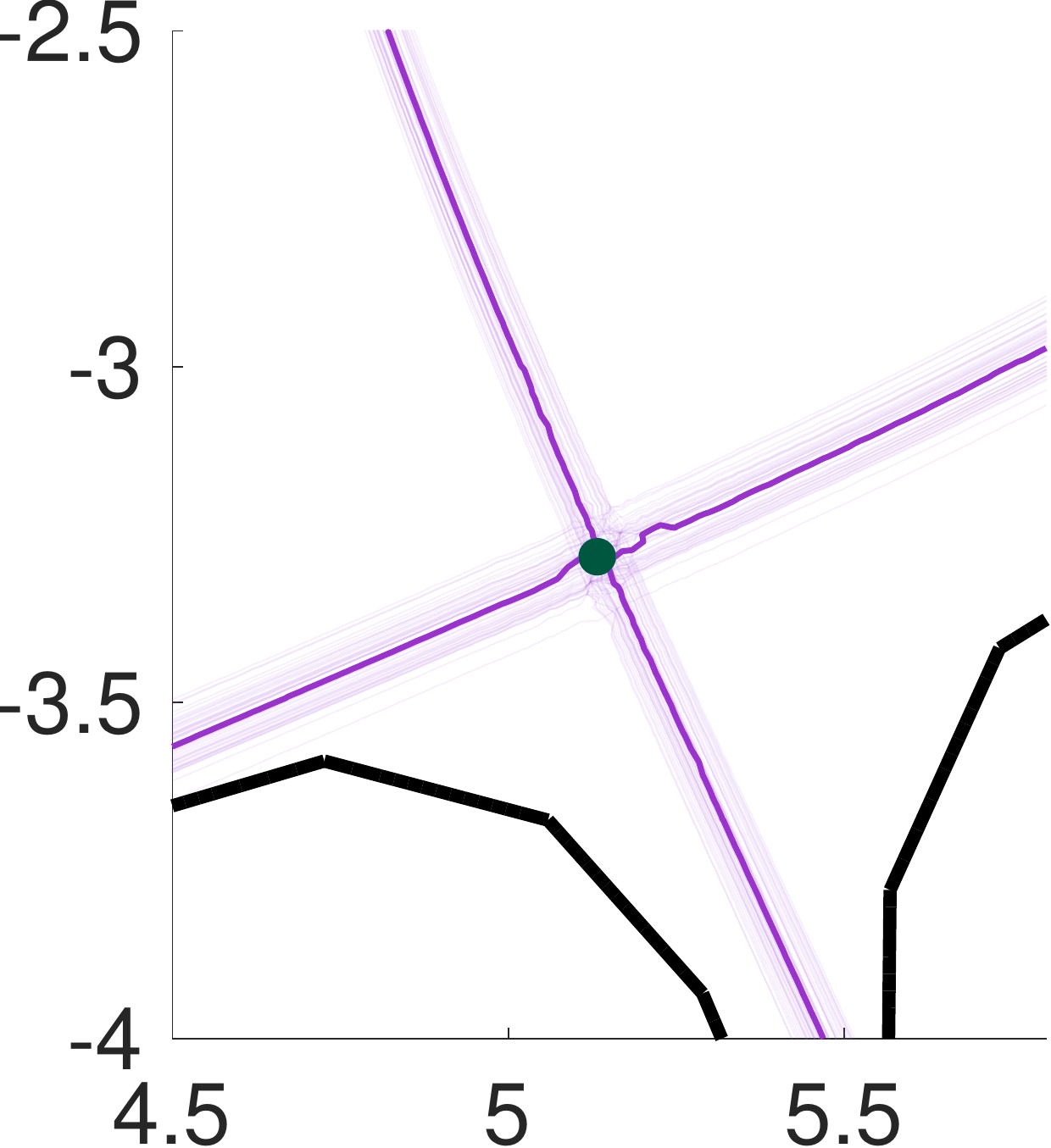} \\[1ex]
\quad MC simulation &\qquad & \quad Uniform MLMC  &\qquad & \quad Adaptive MLMC  \\[-0.5ex]
\quad {\footnotesize Geometry-conforming mesh}&\qquad & \quad {\footnotesize Geometry-conforming mesh} &\qquad & \quad \\
\end{tabular}
\caption{The overlayed plasma boundaries of 50 random realizations are displayed in the top row as violet curves. The solid violet line is the plasma boundary of the expected poloidal flux generated with tolerance $\epsilon=4\times 10^{-4}$. The inner and outer walls of the reactor are displayed in solid black and dark red respectively. The bottom row shows the regions close to the x-points in more detail. The dark green dots are the x-points of the expected solution. Each column from left to right corresponds to: simulation with the Monte Carlo approach, MLMC simulation on geometry-conforming uniform meshes, and adaptive MLMC simulation. All simulations were performed using the discretization level $\ell=5$.} 
\label{fig:QoI_plot}
\end{figure}

\begin{figure}[ht!]\centering
\begin{tabular}{cccccc}
\!\!\rotatebox[origin=l]{90}{{\footnotesize \qquad Uniform MLMC}}\!\! & \includegraphics[width=0.17\linewidth]{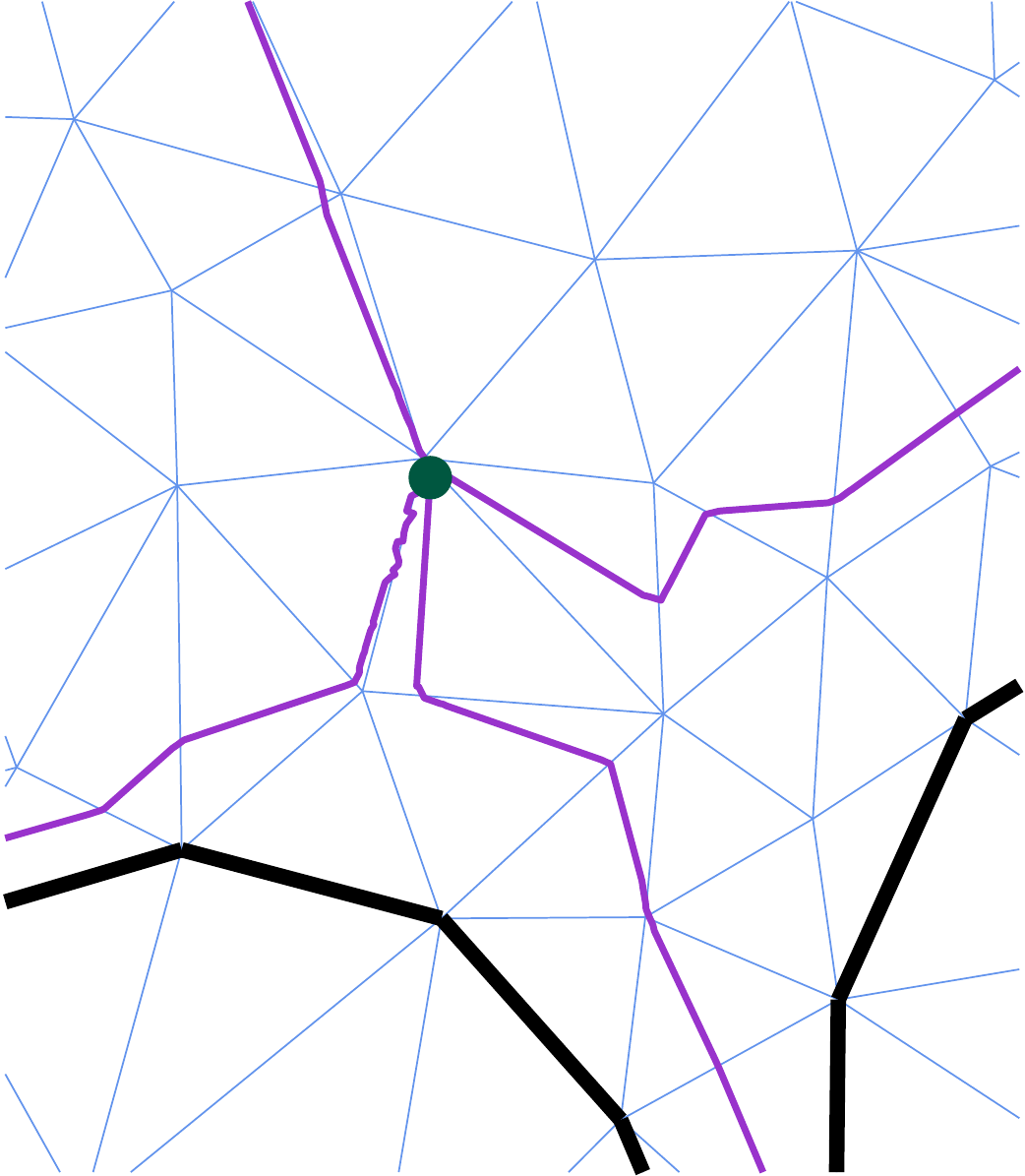} &
\includegraphics[width=0.17\linewidth]{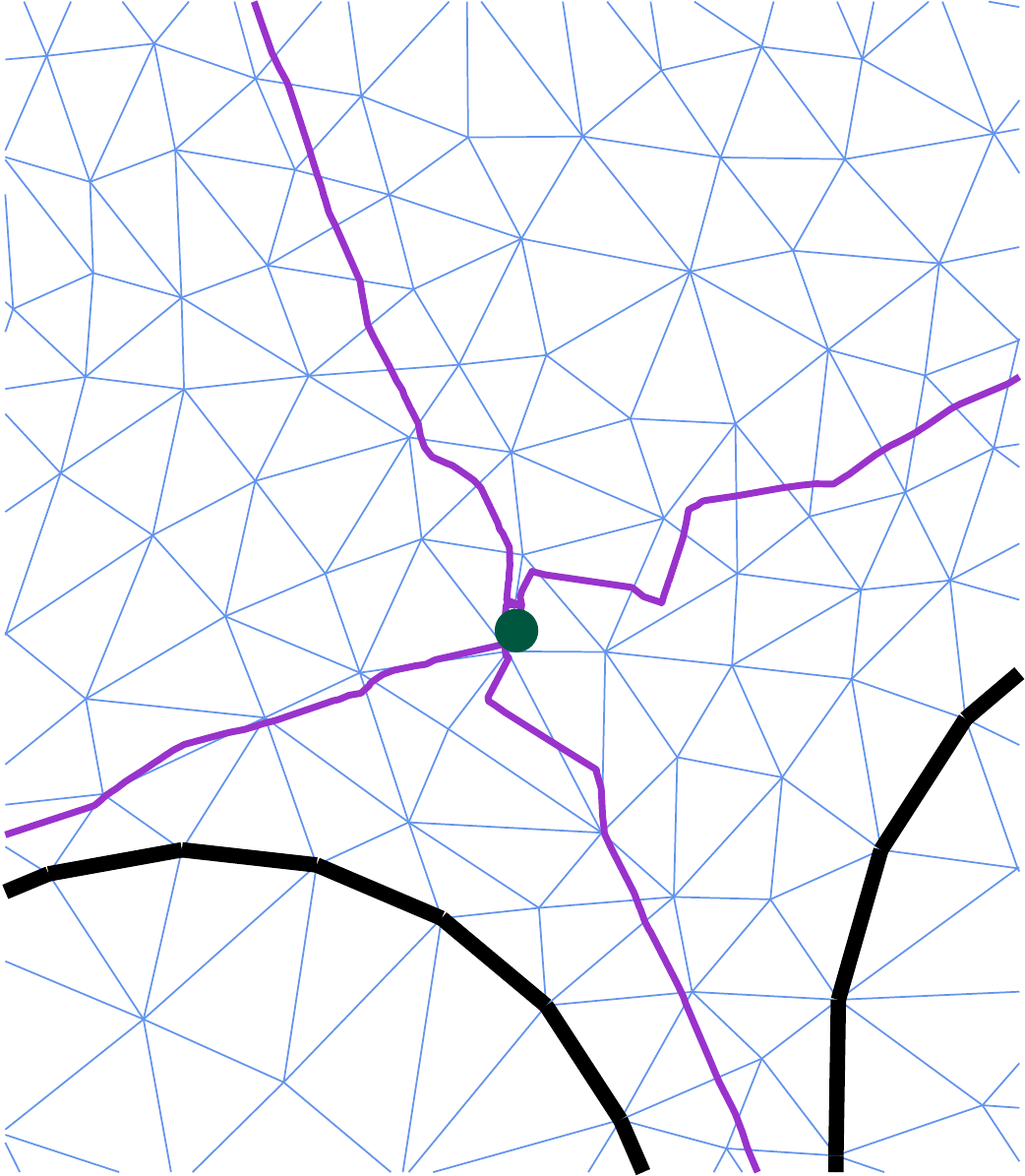} &
\includegraphics[width=0.17\linewidth]{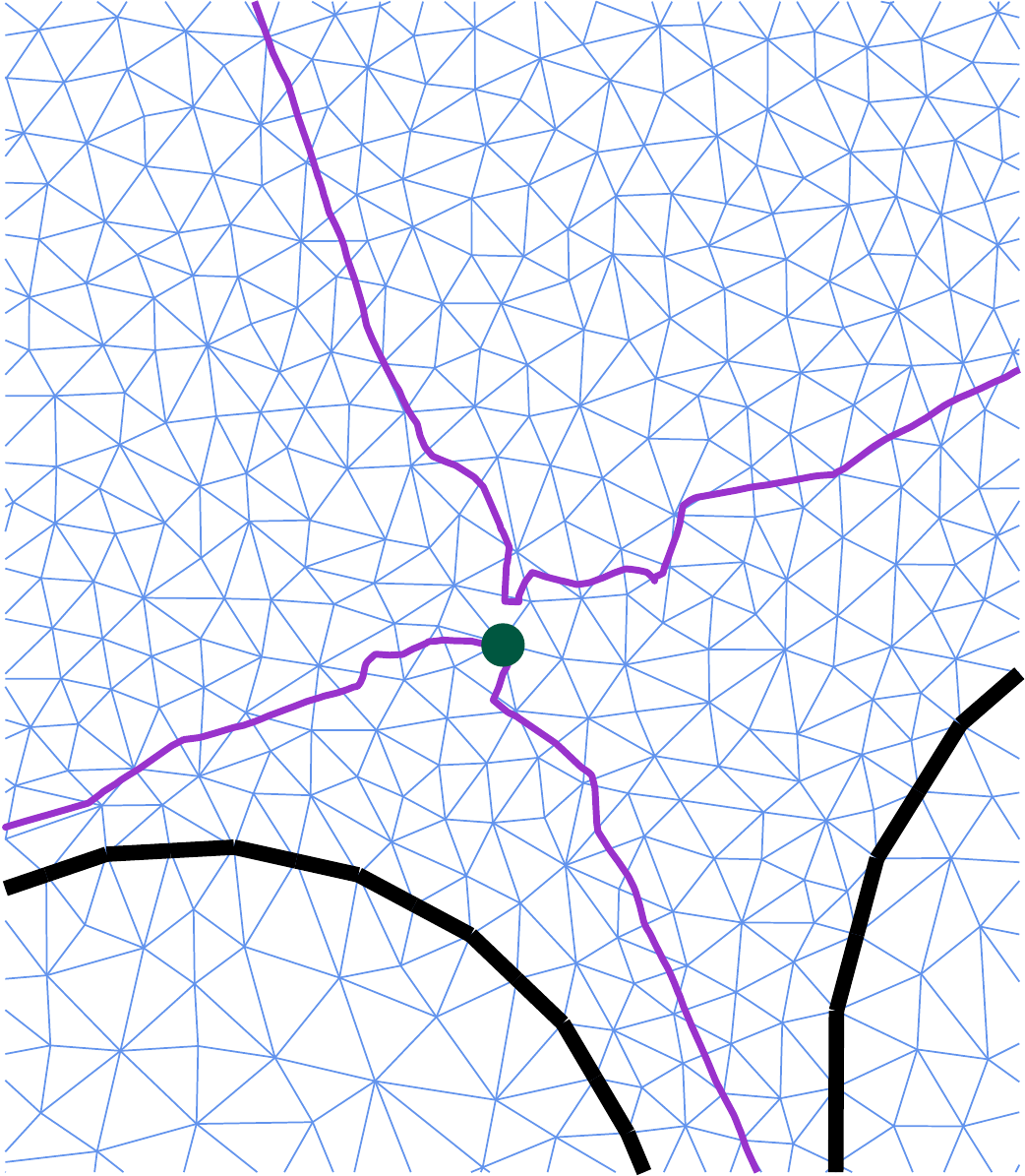} &
\includegraphics[width=0.17\linewidth]{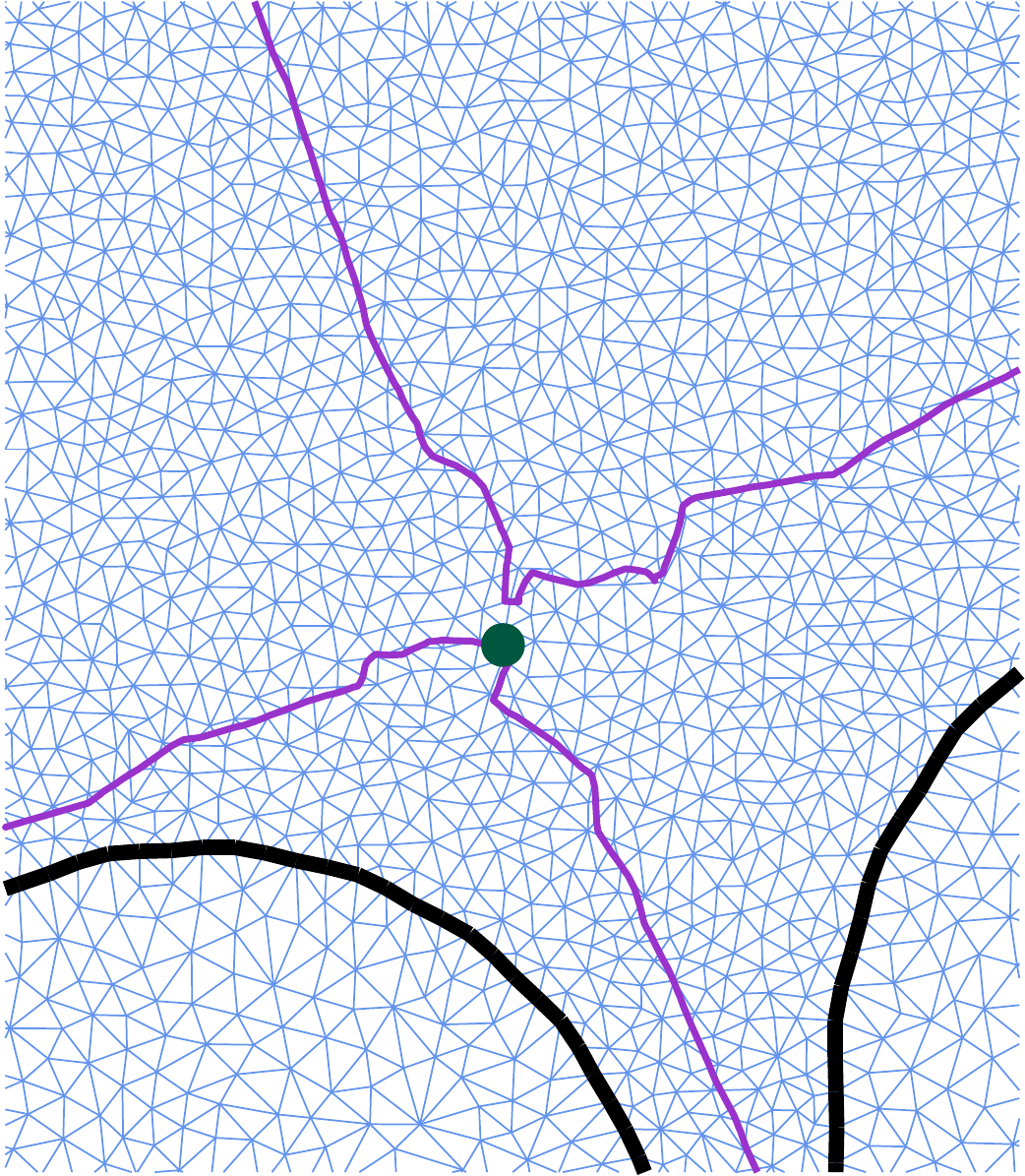} &
\includegraphics[width=0.17\linewidth]{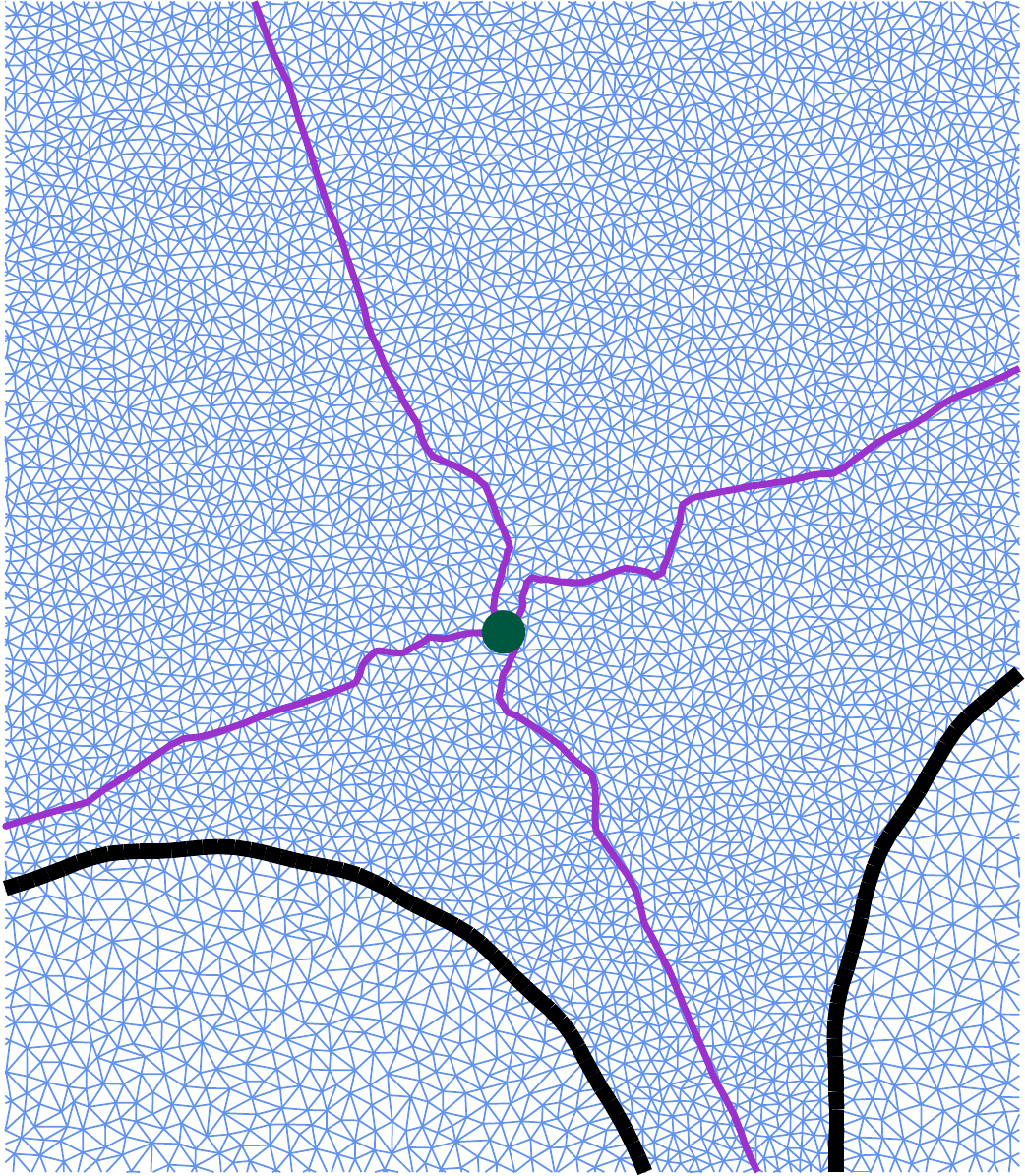} \\[1ex]
\!\!\rotatebox[origin=l]{90}{{\footnotesize \qquad Adaptive MLMC}}\!\! & \includegraphics[width=0.17\linewidth]{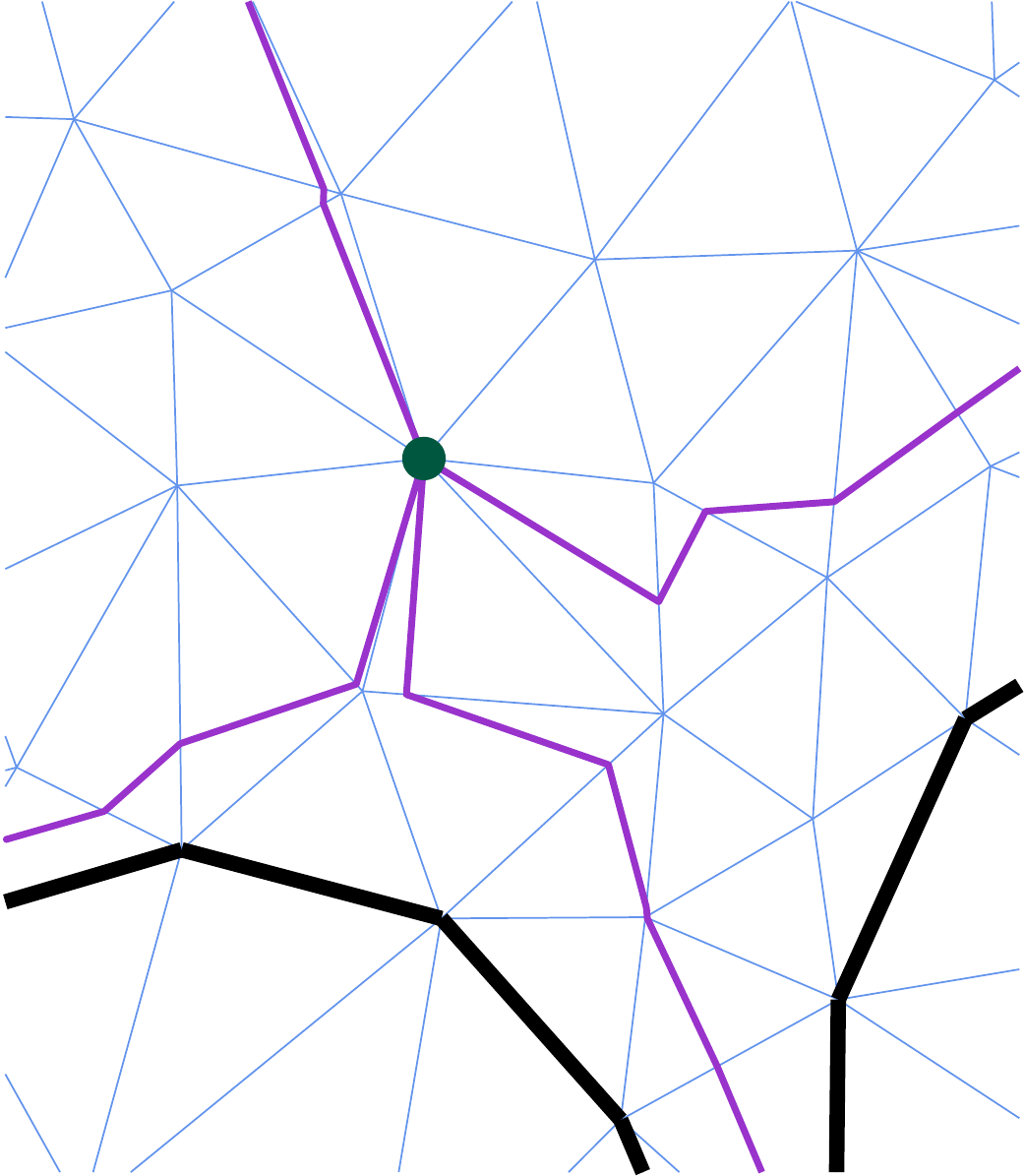} &
\includegraphics[width=0.17\linewidth]{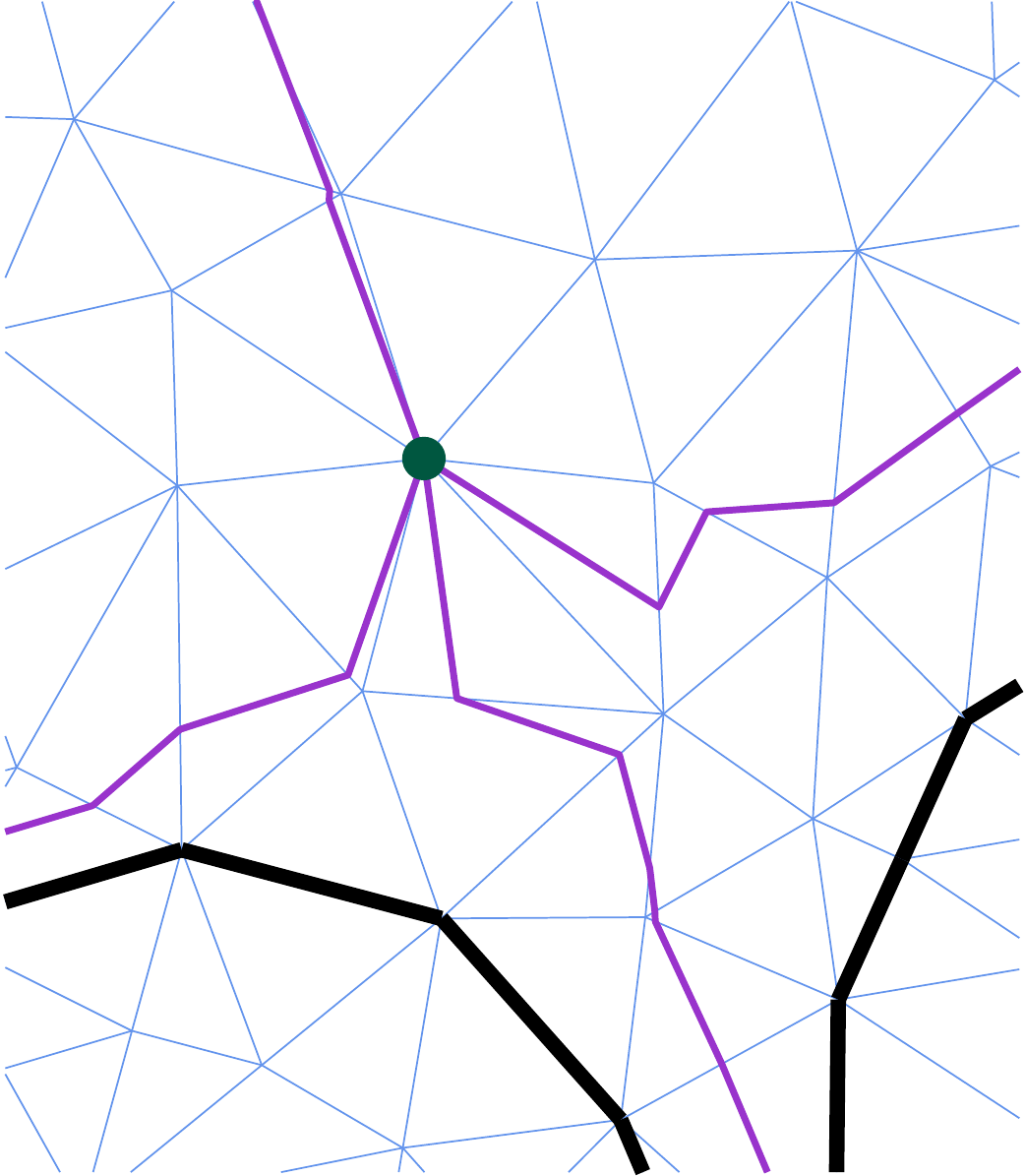} &
\includegraphics[width=0.17\linewidth]{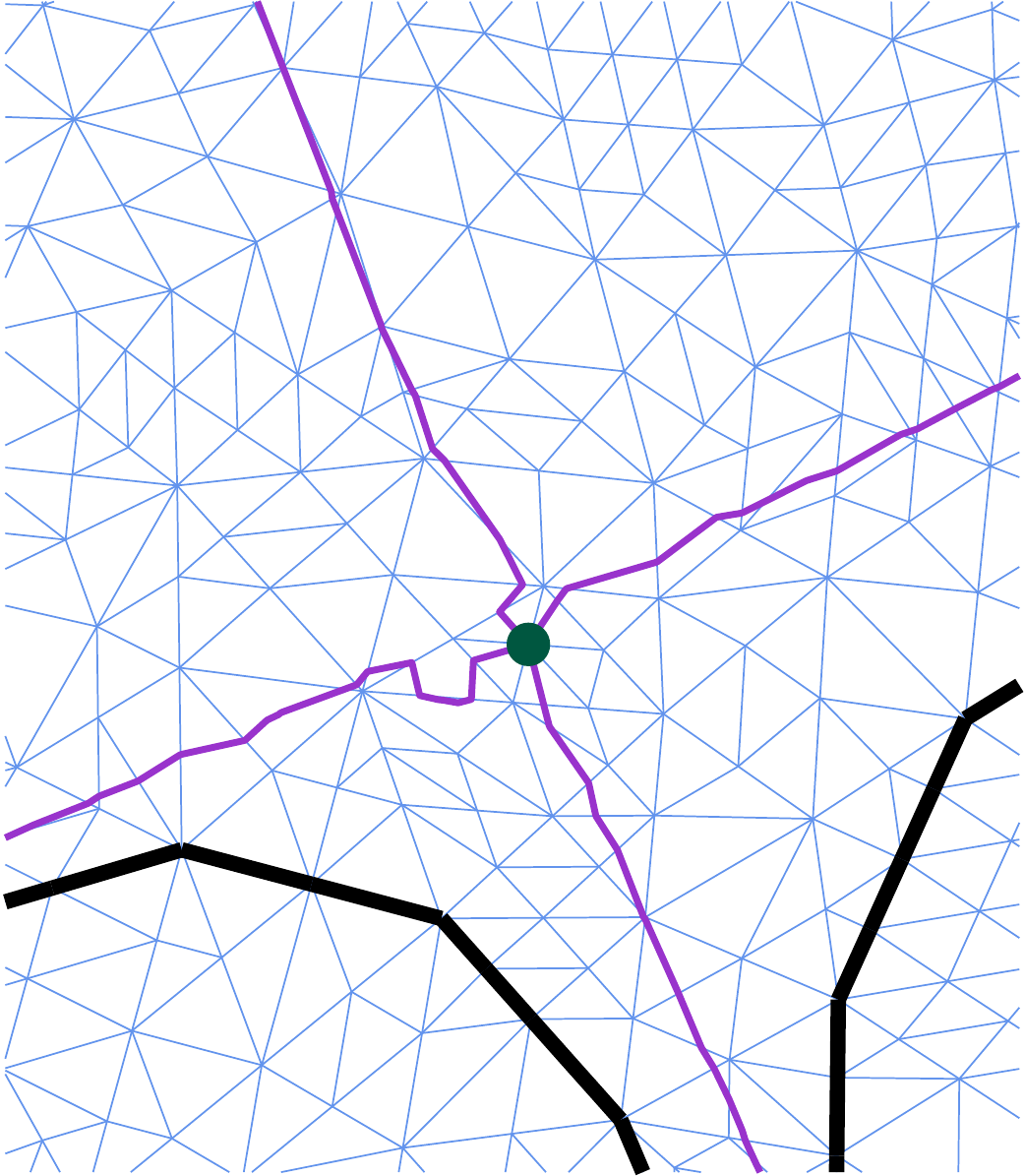} &
\includegraphics[width=0.17\linewidth]{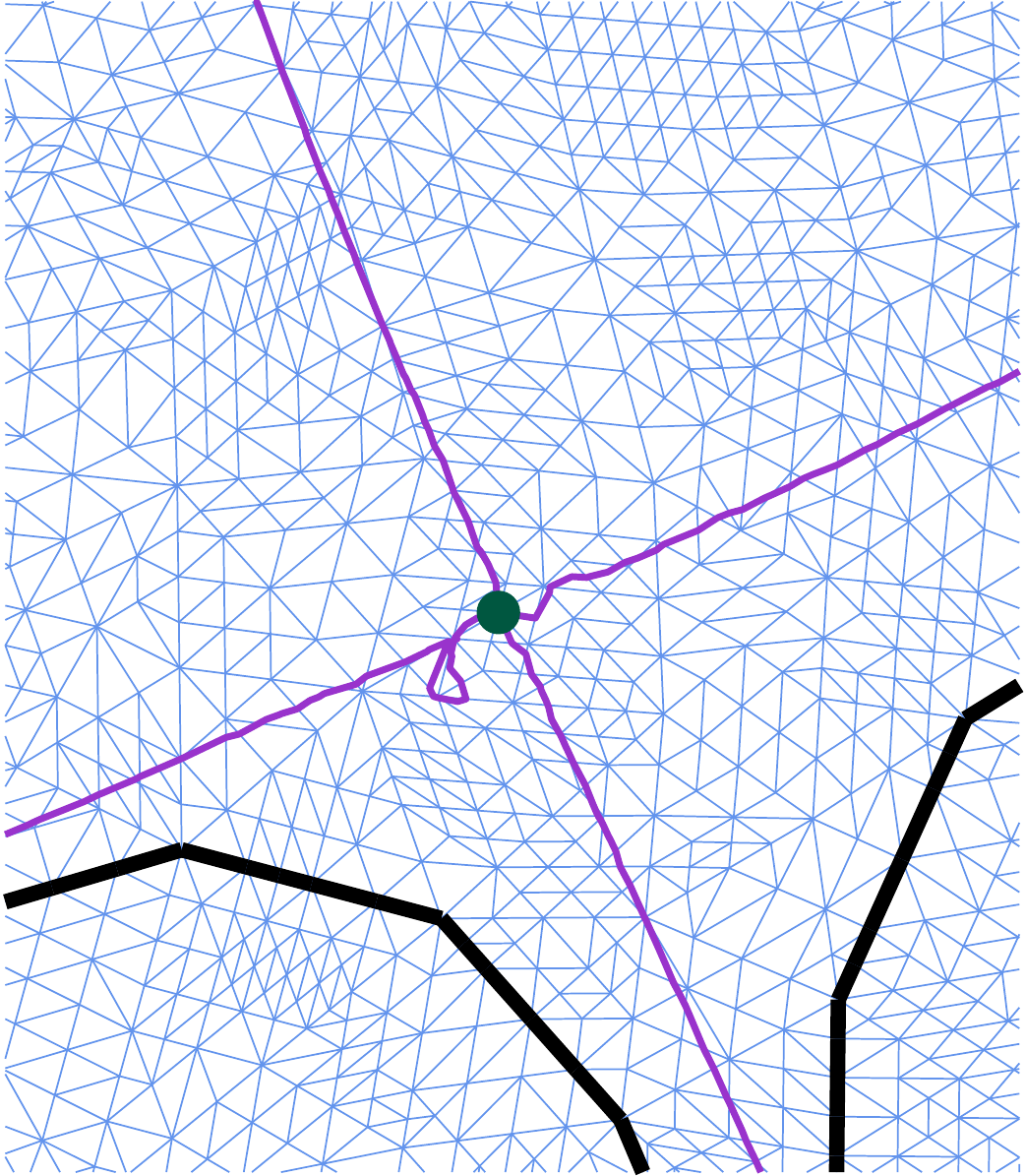} &
\includegraphics[width=0.17\linewidth]{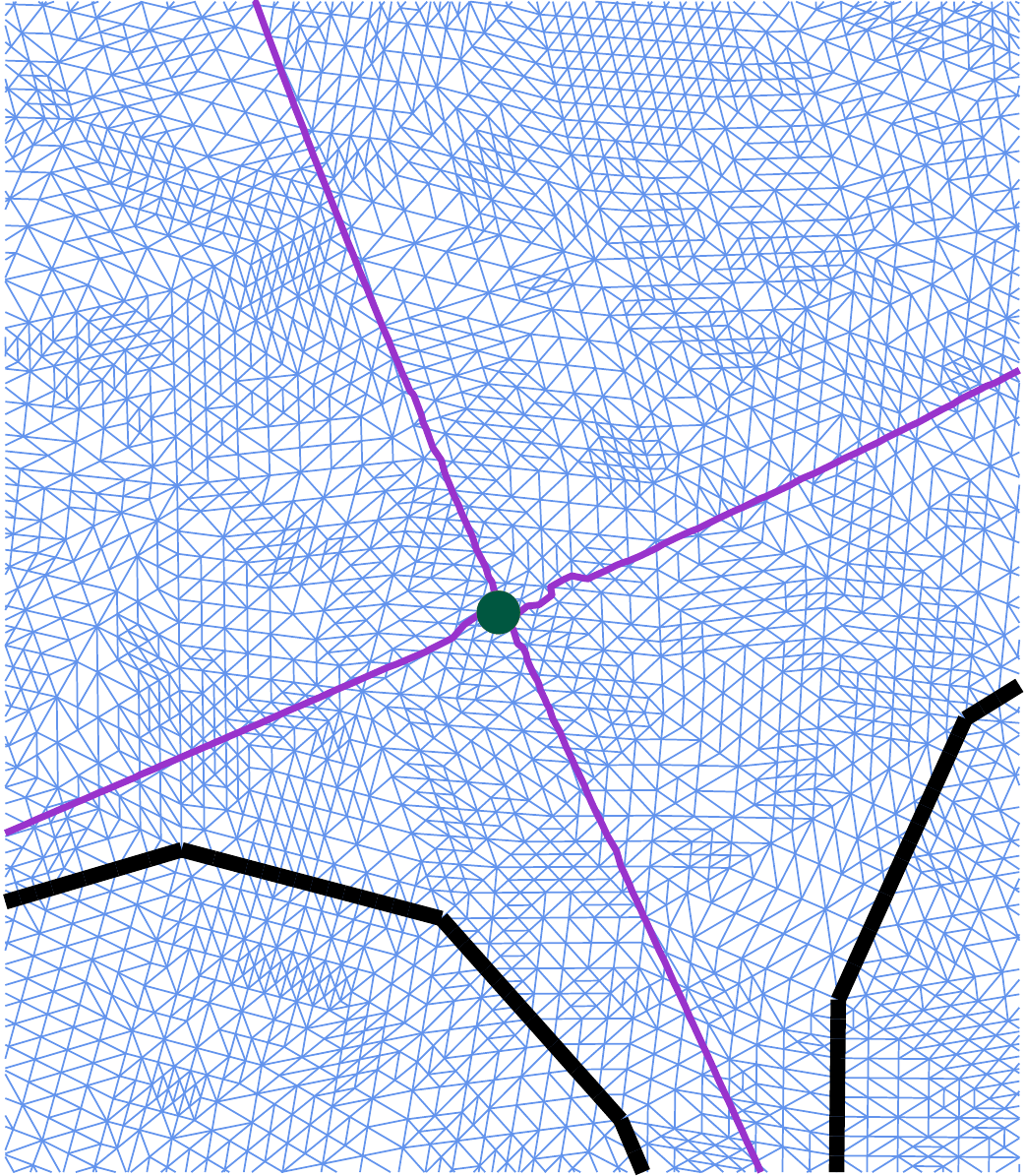} \\
& \quad $\ell=0$ & \quad $\ell=1$  & \quad $\ell=2$ & \quad $\ell=3$& \quad $\ell=4 $\\
\end{tabular}
\caption{The violet curves represent the expected plasma boundaries for simulations using increasingly finer grids when $\epsilon=4\times 10^{-4}$ using the sample sizes specified in Table \ref{Tab:SampleSize}. Each sub-plot focuses on a region near the x-point, maintaining the same zoom-in ratio as the second row of Figure \ref{fig:QoI_plot}. The dark green dots denote the locations of the x-point. The top row shows the results of MLMC-FE on a set of geometry-conforming uniform meshes, while the bottom row displays the results for adaptive MLMC. } 
\label{fig:xpt_CoarseMesh_uniform}
\end{figure}

 \noindent \textbf{Plasma boundary.} To ascertain the expected location of the plasma boundary, we first determine the expected solution to the free boundary problem, $\mathbb E[\psi]$, and determine the boundary of this deterministic function. This boundary is depicted, in dark violet, in Figure \ref{fig:QoI_plot} along with the plasma boundaries obtained from 50 random currents, which are shown in light violet curves. In Figure \ref{fig:xpt_CoarseMesh_uniform} we present plots depicting the x-points and plasma boundaries of the expected solution $\psi$ computed using only samples and corrections from increasingly finer grids for both uniform and adaptive MLMC-FE approaches. The data was obtained with tolerance $\epsilon=4\times 10^{-4}$. As can be seen when moving from left to right in Figure \ref{fig:xpt_CoarseMesh_uniform}, the result obtained using the information from the coarsest level (leftmost column) is progressively corrected with information from increasingly finer grids, leading to the desired result depicted in the rightmost column.

Among the three methods, MC-FE yields the smoothest plasma boundaries in the vicinity of the x-point, followed by adaptive MLMC-FE, while the  MLMC-FE approach on geometry-conforming uniform meshes manifests the most pronounced irregularity in the plasma boundary. The boundary of the expected solution generated with the uniform grid MLMC-FE method exhibits irregularities as can be seen in Figure \ref{fig:QoI_plot}. These large deformations can be primarily attributed to the additional challenges arising from the curved boundaries. We will address this point in more detail at the end of this section. The top row of Figure \ref{fig:xpt_CoarseMesh_uniform} demonstrates that using a geometry-conforming mesh provides a more accurate approximation of the curved structure (in black) of the configuration than that in the bottom row. These observations underscore the challenge of striking a balance between preserving geometric fidelity when dealing with curved boundaries and the desired statistical accuracy of the solution.

\noindent \textbf{Geometric descriptors.}
Table \ref{Tab: QoI_GeoInfo} reports some geometric parameters of the expected poloidal flux $\mathbb E[\psi]$ in \eqref{eq:QoI}. It is observed that these parameters are consistent across different simulation techniques, with agreement typically up to two or, in some cases, one significant digit. Having been derived from $\mathbb E[\psi]$ all these values are deterministic. There is, however, uncertainty associated with the corresponding quantities derived from $\psi$, as they are themselves random variables. Uncertainty on these quantities can be assessed by computing Monte Carlo estimates of their expectations and variances. The multi-level methodology can be used for this purpose without any modification by simply regarding them as quantities of interest in their own right. All the relevant descriptors are computed for each of the samples gathered during the computational loop at the various discretization levels and their expectations and variances are estimated following the same process described on Algorithm \ref{algo:MLMC_Algo_CorrectionVersion}. The expectations and variances obtained in this fashion are shown in Table \ref{Tab:GeoInfo_mean}.

Despite the advantages of low computational cost, the MLMC-FE-based methods may encounter difficulties in accurately determining the locations of x-points and magnetic axis. Note that the x-points, which correspond to saddle points of the piecewise linear approximation of $\psi$, can only be located at the nodes of the mesh. The numerical identification of their exact locations, which often relies on changes in the sign of the discrete gradient, can be challenging; see \cite{BiClCh:2016,ElLiSa:2022, Hofmann1988} for discussions of the computational difficulties.

In summary, simulations using the uniform MLMC-FE on non-nested geometry-conforming uniform meshes may encounter a substantial challenge in accurately identifying the x-point and achieving less accurate quantities, especially for the plasma boundary, when compared to the results obtained from MC-FE. On the other hand, the adaptive MLMC-FE approach on a nested adaptively refined mesh set produces results that closely align with the MC-FE at a much lower computational cost.

\begin{table}[ht]
	\centering
			\scalebox{0.6}{
		\begin{tabular}{c|c|c|c|c|c|c|c|c|}
			\cline{2-4}
				&\multicolumn{1}{c|}{MC-FE}&Uniform MLMC-FE&Adaptive MLMC-FE\\
			\hline
			\multicolumn{1}{|c|}{x point}&(5.14,-3.29)&(5.14,-3.29)&(5.14,-3.28)\\
			\hline
			\multicolumn{1}{|c|}{magnetic axis}&(6.41,0.61)&(6.44,0.56)&(6.46,0.54)\\
			\hline
			\multicolumn{1}{|c|}{strike} &(4.16,-3.71)&(4.16,-3.71)&(4.16,-3.71)\\
			\multicolumn{1}{|c|}{points}&(5.56,-4.22)&(5.56,-4.22)&(5.56,-4.21)\\
			\hline
			\multicolumn{1}{|c|}{inverse aspect ratio} &0.32&0.32&0.32\\
			\hline
			\multicolumn{1}{|c|}{elongation} &1.86&1.87&1.86\\
			\hline
			\multicolumn{1}{|c|}{upper triangularity}&0.43&0.43&0.43\\
			\hline
			\multicolumn{1}{|c|}{lower triangularity} &0.53&0.53&0.53\\
			\hline
	\end{tabular}
  }
	\caption{Geometric parameters of the expected poloidal flux $\mathbb E[\psi]$ from MC-FE, MLMC-FE with geometry-conforming uniform mesh set, and adaptive MLMC-FE. The results are generated with an nMSE $4\times 10^{-4}$.}
	\label{Tab: QoI_GeoInfo}
\end{table}

\begin{table}[ht]
	\centering
			\scalebox{0.65}{
		\begin{tabular}{c|c|c|c|c|c|c|c|c|}
			\cline{2-7}
				&\multicolumn{2}{c|}{MC}&\multicolumn{4}{c|}{MLMC}\\
			\cline{2-7}
			\multicolumn{1}{c|}{}&\multicolumn{2}{|c|}{-}&\multicolumn{2}{|c|}{Uniform}&\multicolumn{2}{|c|}{Adaptive}\\
   			\hline
			\multicolumn{1}{|c|}{Geometric parameters}&Mean &Variance&Mean &Variance&Mean &Variance\\
			\hline
			\multicolumn{1}{|c|}{x point}&(5.14,-3.29)&(2.37e-04,1.44e-03)&(5.14,-3.28)&(2.44e-04,6.33e-04)&(5.14,-3.29)&(4.03e-04,2.91e-03)\\
			\hline
			\multicolumn{1}{|c|}{magnetic axis}&(6.41,0.61)&(1.01e-03,6.11e-04)&(6.41,0.60)&(2.65e-02,6.28e-03)&(6.42,0.60)&(3.15e-02,5.81e-03)\\
			\hline
			\multicolumn{1}{|c|}{strike} &(4.17,-3.71)&(4.46e-03,2.23e-03)&(4.16,-3.71)&(1.23e-04,4.30e-05)&(4.17,-3.71)&(4.18e-04,2.00e-04)\\
			\multicolumn{1}{|c|}{points}&(5.56,-4.22)&(2.88e-08,3.28e-03)&(5.56,-4.22)&(4.05e-09,5.54e-05)&(5.56,-4.21)&(2.02e-08,2.98e-04)\\
			\hline
			\multicolumn{1}{|c|}{inverse aspect ratio} &0.32&4.76e-06&0.32&4.23e-06&0.32&4.31e-06\\
			\hline
			\multicolumn{1}{|c|}{elongation} &1.86&1.50e-04&1.86&1.83e-03&1.86&2.29e-03\\
			\hline
			\multicolumn{1}{|c|}{upper triangularity}&0.43&2.49e-04&0.43&1.45e-03&0.43&2.08e-03\\
			\hline
			\multicolumn{1}{|c|}{lower triangularity} &0.53&1.08e-04&0.53&1.52e-03&0.53&1.31e-03\\
			\hline
	\end{tabular}}
	\caption{Sample mean of the geometric parameters extracted from each realization during three simulations MC-FE, uniform MLMC-FE, and adaptive MLMC-FE. The results are generated with an nMSE $4\times 10^{-4}$.}
	\label{Tab:GeoInfo_mean}
\end{table}

\noindent \textbf{Meshing curved domains and their effect on MLMC estimations.} In the deterministic setting this \textit{geometric error} has the undesired consequence of hindering the decay of the discretization error since, as the mesh is refined, the discretized computational domain does not converge to the semicircle bounded by $\Gamma$. In the stochastic setting the geometric error manifests itself in rendering the Monte Carlo estimator biased and inconsistent. The inconsistency stems from the fact that, as both the sample size and the mesh level increase, the Monte Carlo estimator does not converge to the expectation of the random variable $u$ satisfying the free boundary problem. Instead, the estimator converges to the expectation of the random variable that satisfies a perturbation of \eqref{eq:VariationalFormulation} where the curve $\Gamma$ is not a semicircle, but the initial polygonal approximation. If the initial mesh is fine enough, this geometric bias will likely be too small to affect the estimation.

On the other hand, if an exact descriptor of the curved boundaries is available, the aforementioned difficulty can be overcome by re-sampling the curved boundaries when building the sequence of finer grids, thus allowing for a resolution of the curved structures consistent with the respective mesh parameter. If an exact descriptor is not available it is possible to approximate it with, for instance, a cubic spline representation that interpolates the original polygonal representation. This spline surrogate is then used to re-sample the boundary as the mesh is refined. This strategy, which gives rise to what we refer to as \textit{geometry conforming meshes}, was implemented for the numerical experiments with \textit{uniformly refined meshes} and can be seen in use in the top row of Figure \ref{fig:xpt_CoarseMesh_uniform}, where the curved boundary is represented more accurately as the mesh is refined. 

\begin{figure}[ht!]\centering
\begin{tabular}{ccccc}
\includegraphics[width=0.18\linewidth]{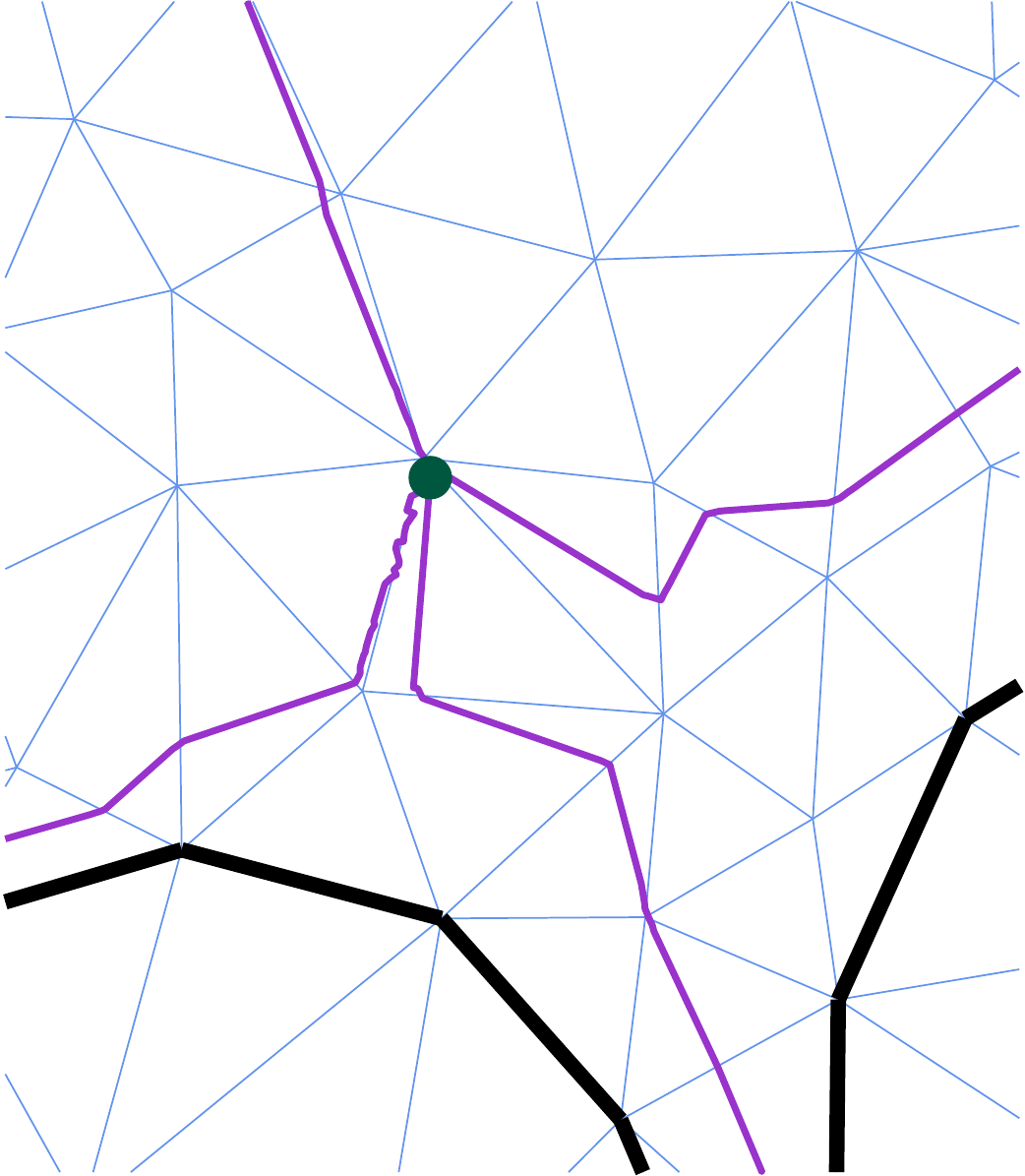} &
\includegraphics[width=0.18\linewidth]{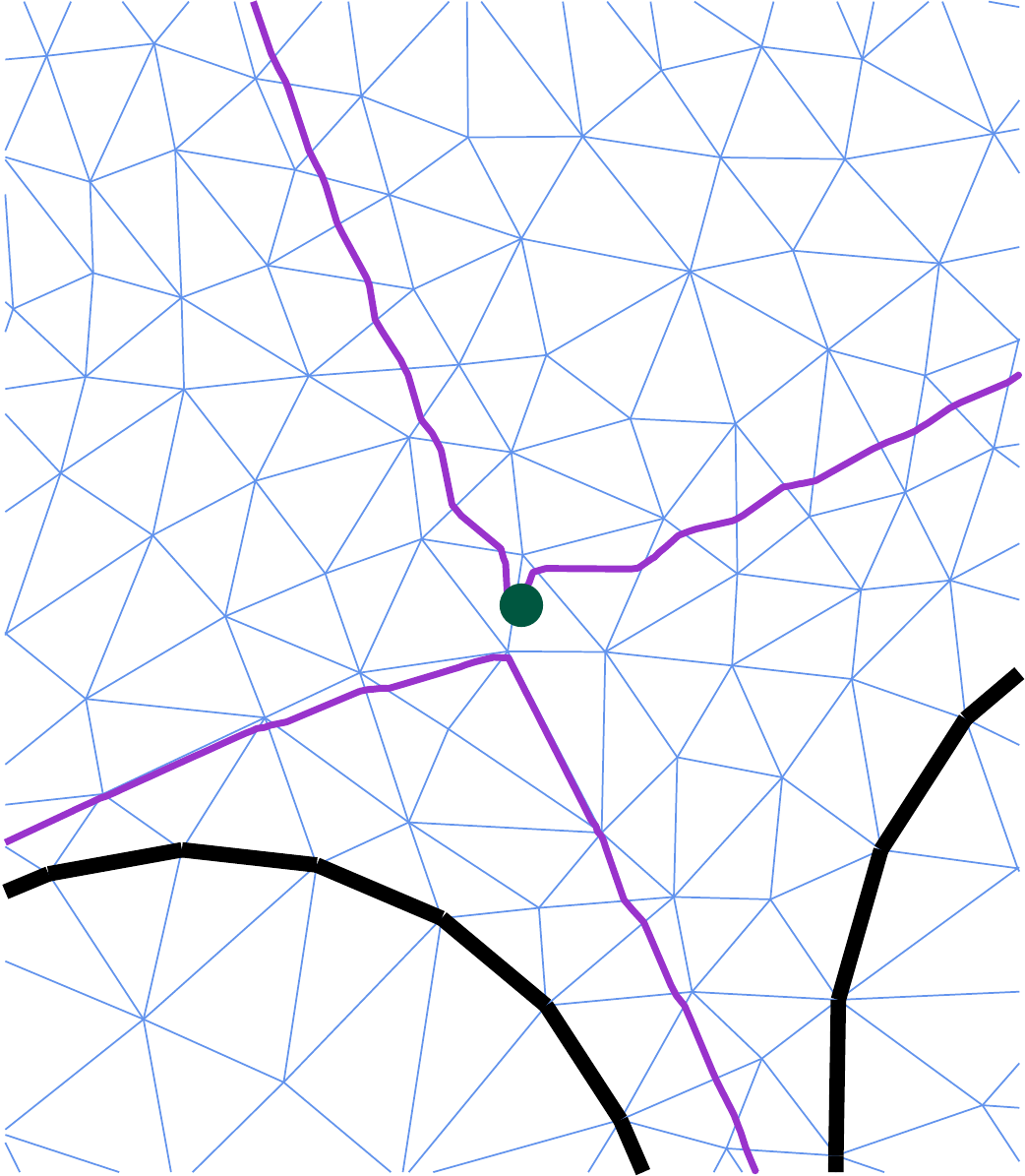} &
\includegraphics[width=0.18\linewidth]{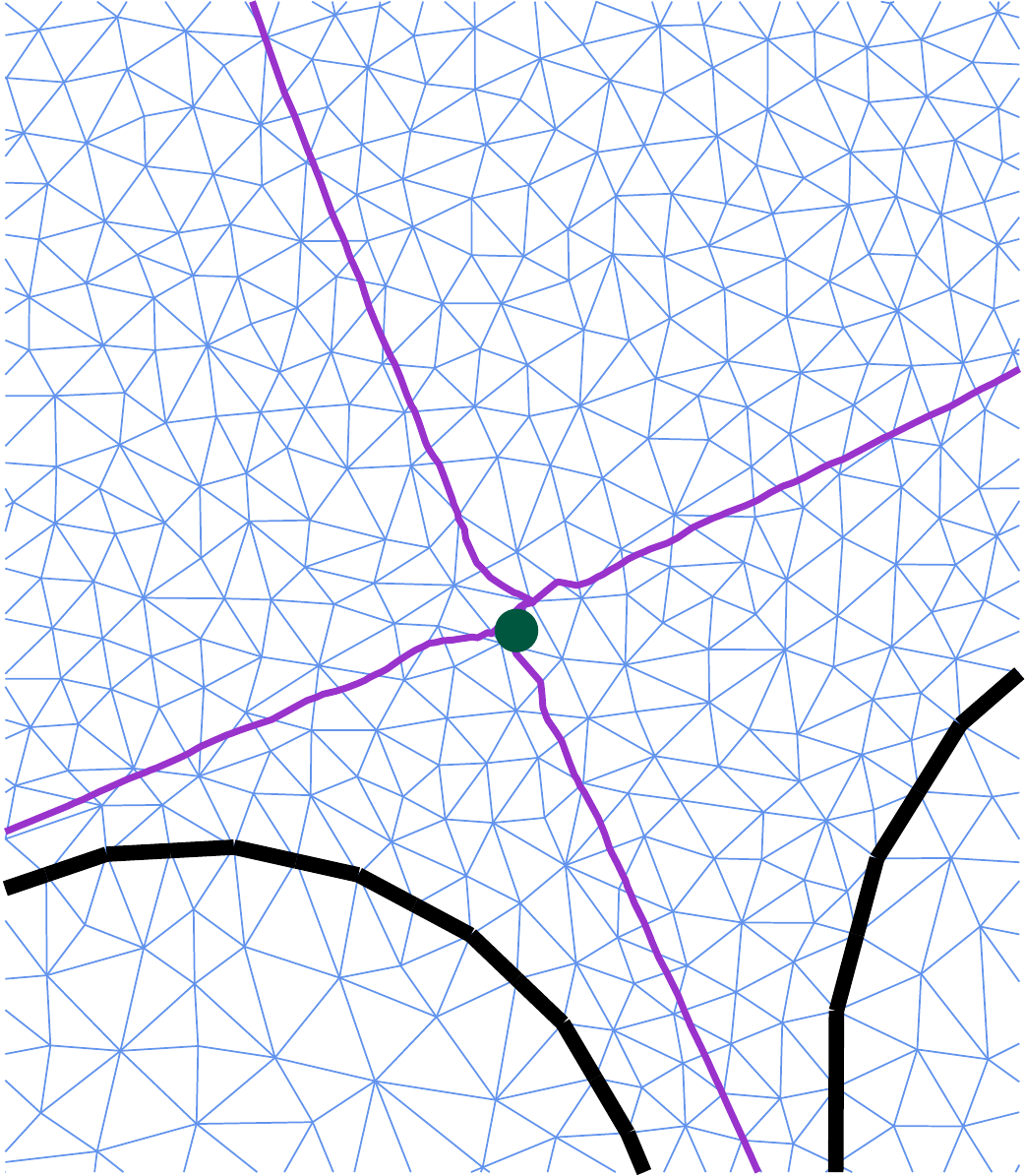} &
\includegraphics[width=0.18\linewidth]{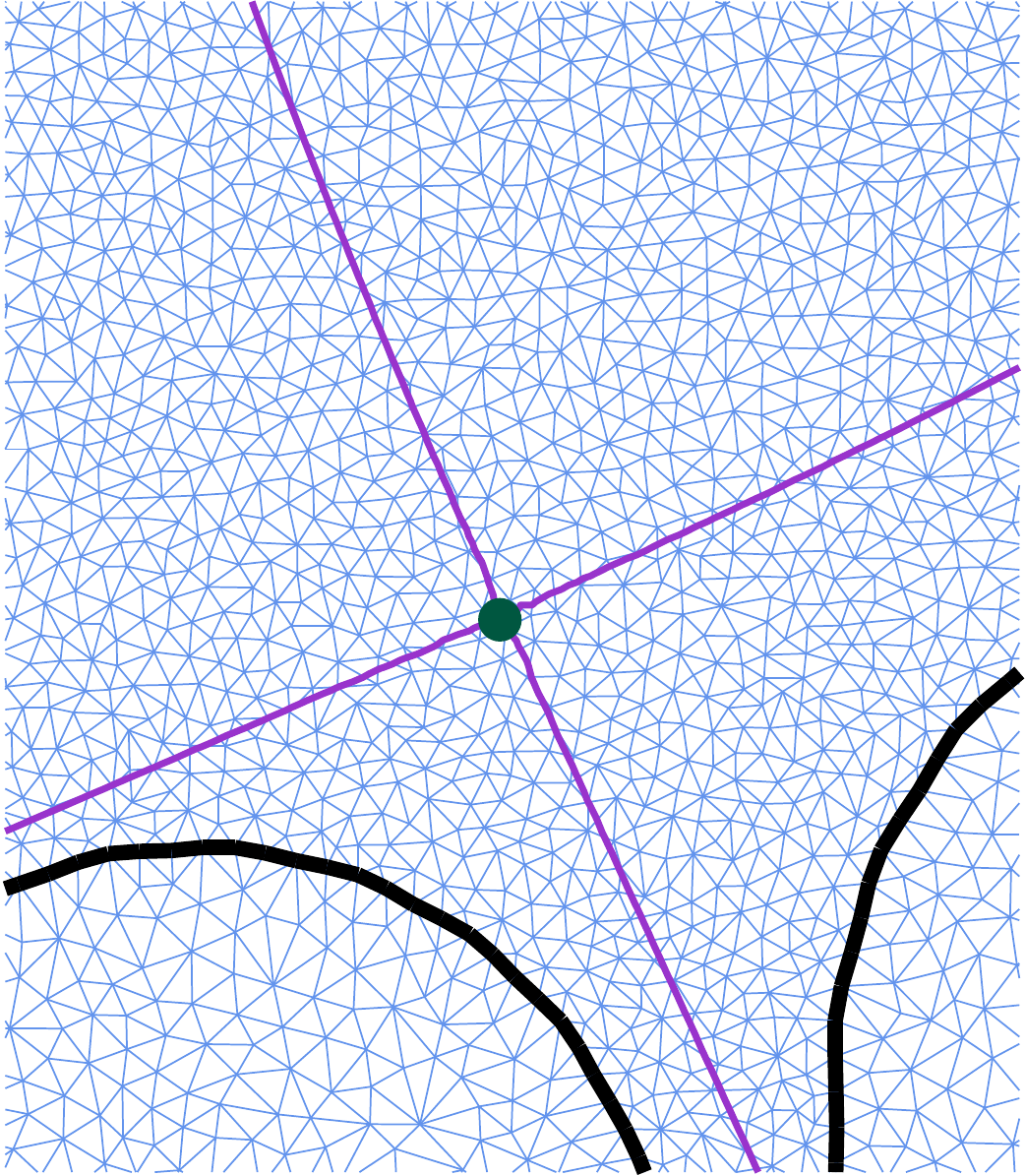} &
\includegraphics[width=0.18\linewidth]{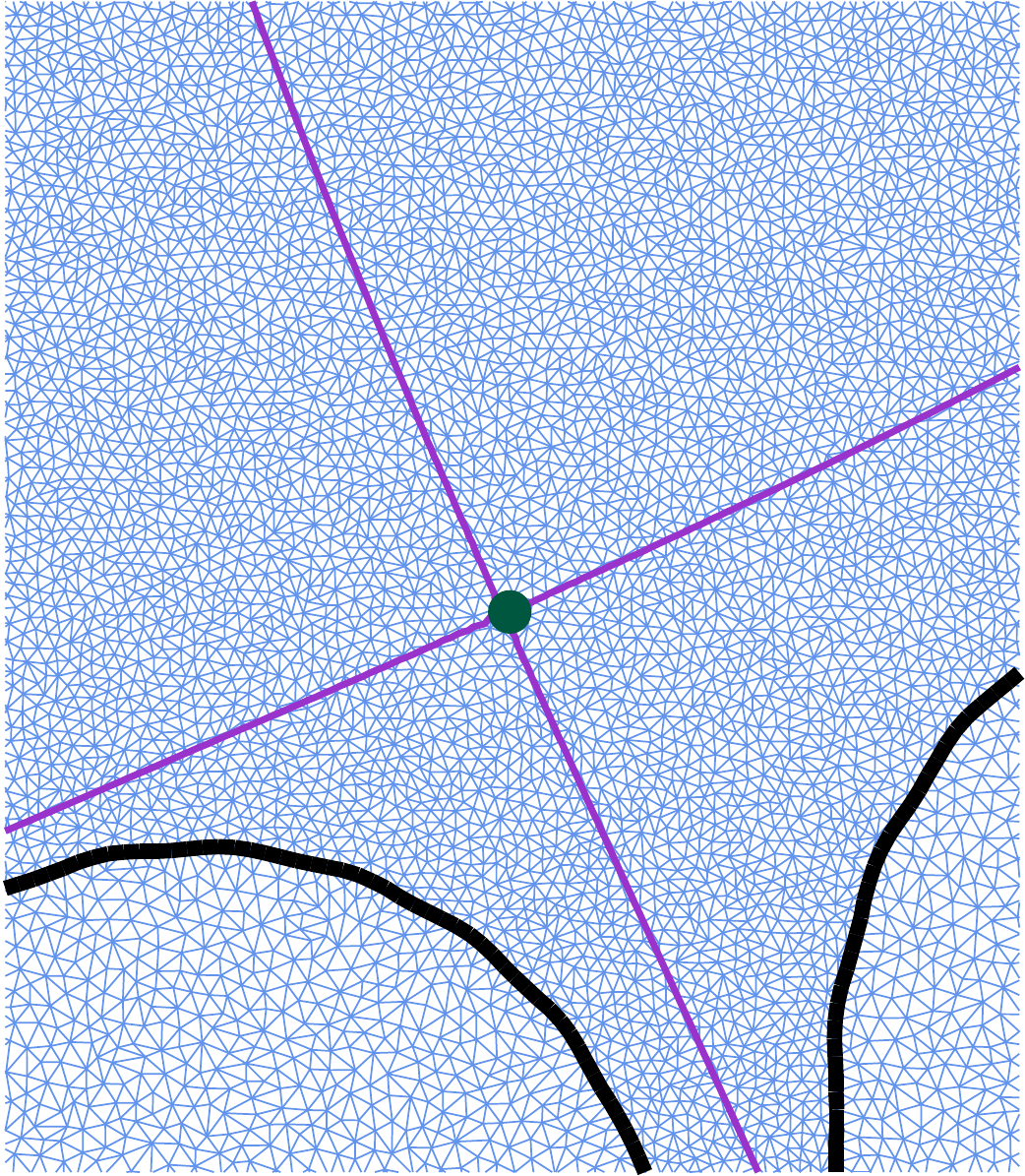}\\
\quad $\ell=0$ & \quad $\ell=1$  & \quad $\ell=2$ & \quad $\ell=3$& \quad $\ell=4$\\
\end{tabular}
\caption{The violet curves represent the expected plasma boundaries for post-processed simulations using increasingly finer grids when $\epsilon=4\times 10^{-4}$ and sample size specified in Table \ref{Tab:SampleSize}. Each sub-plot focuses on a region near the x-point, maintaining the same zoom-in ratio as the second row of Figure \ref{fig:QoI_plot}. The dark green dots denote the locations of the x-point.} 
\label{fig:xpt_CoarseMesh_uniform_no_wiggles}
\end{figure}

Nevertheless, even if the approximation to the problem geometry is now consistent with the discretization error, this approach creates additional challenges. Since the approximations to the curved boundaries are not fixed across mesh levels, the sequence of meshes is no longer nested---not even in the case of uniform refinements. Moreover, due to the fact that the sequence of discrete domains no longer coincides across levels, the domains of definition of the respective discrete solutions will not overlap, and an extrapolation step may be needed to compute the multilevel Monte Carlo estimator on a common computational domain. This strategy, used in our numerical experiments, introduces an additional extrapolation error. In our case this is evident, for instance, in the fact that the plasma boundary of the expected solution $\mathbb E[\psi_h]$ is considerably less regular in the geometry-conforming case than it is in the non-geometry-conforming one. This can be seen in Figure \ref{fig:xpt_CoarseMesh_uniform}. The extrapolation error can be taken care of through careful post-processing. One option is to project or interpolate the numerical solutions into a subdomain common to all grids so that no extrapolation is needed for evaluation. This strategy was employed to produce Figure \ref{fig:xpt_CoarseMesh_uniform_no_wiggles} successfully eliminating the spurious oscillations in the plasma boundary. However, doing this requires considerable computational work and reduces the time savings obtained from MLMC.

One further difficulty is that the re-sampling of the boundaries is impossible to perform in a straightforward fashion in the case of adaptively refined meshes. Thus, the geometric approximation remains fixed at the initial level of refinement. This can be seen in the bottom row of Figure \ref{fig:xpt_CoarseMesh_uniform}, where the solid black line represents the polygonal approximation to the curved boundary of the divertor. The approximation improves as the mesh is refined for the uniformly refined mesh, but stays fixed for the adaptive strategy.  

\section{Concluding remarks}
\label{sec:Conclusion}
The objective of this study is to evaluate the performance of MC-FE and several variants of  MLMC-FE for the Grad-Shafranov free boundary problem under high-dimensional uncertainties in the currents. The plasma equilibrium problem presents challenges and complexities associated with the physical properties of the system, the values of the parameters of the experiment, and even the approximation of geometric features of the model. Examples include the degradation of the stability of the equilibrium associated with an increase of plasma elongation or beta; the presence of finite current near the plasma boundary or, as in this study, the complex boundary structures, which must be treated carefully in the context of multilevel methods. These physically grounded challenges typically get reflected in the computations, for instance, as an increase in the difficulty of a linearized algorithm to converge to a desired tolerance or as the need for higher precision in the numerical computation required to resolve the desired features. In other words, complicated physical scenarios typically result in an increase in computational costs which might easily become overwhelming when combined with uncertainty quantification efforts. In cases like these, an MLMC strategy can help to mitigate the costs required to gather statistical information at the expense of potentially overlooking subtle behavior that may not be resolved in the samples gathered in coarse levels of refinement.

It is important to remark that the multilevel Monte Carlo techniques presented rely only on the independence between different realizations \textit{across discretization levels} and not on the particular probability distribution obeyed by the stochastic parameters. In this study, the current intensities were modeled as independent uniformly distributed random variables, but the method would have worked without any modification even if the current intensities had obeyed correlated distributions \textit{as long as realizations across discretization levels are independent of each other}. Moreover, further stochastic components---such as the parameters appearing in the profile models in equation \eqref{eq:source}---can be included and handled simultaneously.

In this study, we handled the geometric difficulties in the uniform MLMC-FE case using a sequence of uniform meshes that conform to the curved boundaries in the geometry. We also found that a feature of adaptive MLMC-FE is that adaptive gridding leads to a somewhat better representation of geometric quantities; in contrast, errors introduced by computations on uniform coarse meshes lead to some distortions of some features. The traditional advantage of adaptive refinement reduced computational cost, was less clearly present. But the overall advantage of the MLMC approach in reducing costs is dramatic.

\section{Acknowledgements}
All the free boundary computations were done using the code 
{\tt FEEQS.M} \cite{Heumann:feeqsm,CEDRES}. The authors are deeply grateful to Holger Heumann, the INRIA CASTOR team, and all the development team of the CEDRES++ free boundary solver for sharing access to the code and for helping us get up to speed with its usage.

Howard Elman has been partially supported by the U. S. Department of Energy under grant DE-SC0018149 and by the U. S. National Science Foundation under grant DMS1819115. Jiaxing Liang has been partially funded by the U. S. Department of Energy under grant DE‐SC0018149. Tonatiuh S\'anchez-Vizuet has been partially funded by the U. S. National Science Foundation through the grant NSF-DMS-2137305.

\section{Appendix}\label{sec:Appendix}
We now describe some technical mathematical and implementation details related to the work presented. This additional material is provided for the sake of completeness but is not needed to follow the discussion from the previous sections. 
%
\subsection{Weak formulation of the Grad-Shafranov equation} \label{sec:Space_Norm_VariationalFormulation}
%
\noindent\textit{Deterministic problem.} We will start by introducing the weak formulation for the Grad-Shafranov problem in the case where all the parameters involved are deterministic. Consider a semi-circle centered at the origin, boundary $\Gamma$ and with radius $\rho$ such that it fully contains all the relevant reactor components depicted in Figure \ref{fig:Geometry}. If $\Omega$ denotes the region surrounded by $\Gamma$ then, by construction, for any in $\Omega^c$ the right-hand side of \eqref{eq:FreeBoundarya} will vanish identically. We will then, following \cite{Gr:1999}, consider the space of real-valued functions

\begin{equation}
\label{eq:FunctionSpace}
    Z:=\left\{u:\Omega\rightarrow \mathbb{R} \,\Bigg| \,\int_\Omega u^2x\,dxdy<\infty \,,\, \int_\Omega\frac{\left|\nabla u\right|^2}{x}dxdy<\infty; \,\text{ and  }\,u(0,y)=0 \right\}\cap C^0\left(\overline{\Omega}\right).
\end{equation}

This space arises naturally when testing equation \eqref{eq:FreeBoundarya} using the weighted $L_2$ inner product defined by
\[
    \langle u,v\rangle := \int_{\Omega} u v x\,dxdy,
\]
which leads to the finite energy requirement appearing in the second inequality in the definition \eqref{eq:FunctionSpace}. The third requirement ($u(0,y)=0$) is a result of the anti-symmetry of the problem with respect to reflections across the axis of symmetry of the reactor and has the effect of ensuring that the quantity
\begin{equation}
\label{eq:EnergyNorm}
    \left\Vert u \right\Vert_Z :=\left(\int_\Omega\frac{\left|\nabla u\right|^2}{x}dxdy\right)^{1/2}
\end{equation}
does indeed define a norm in the space $Z$. We will refer to this norm as the \textit{energy norm}. The space $Z$ defined in \eqref{eq:FunctionSpace} endowed with the energy norm \eqref{eq:EnergyNorm} is the natural function space to look for variational solutions to the deterministic linearized Grad-Shafranov equation \eqref{eq:FreeBoundary}. The variational formulation of the problem, as derived in \cite{AlBlDe:1986,GaHs:1995,HsZh:1994}, is that of finding $\psi\in Z$ such that for every test $\varphi\in Z$ it holds that:
\begin{align}
\nonumber
& \int_{\Omega_p}\frac{1}{\mu x}\nabla\psi\cdot\nabla\varphi\,dxdy
-\int_{\Omega_p}\left(rp^\prime(\psi) + \frac{1}{\mu_0 x}ff^\prime(\psi)\right)\varphi \,dxdy+\int_\Gamma \psi\,N\,\varphi\,dc \\ 
\label{eq:VariationalFormulation}
&\hspace{.2in}
 + \int_\Gamma \int_\Gamma  \left(\psi(\boldsymbol x_1) - \psi(\boldsymbol x_2)\right) M(\boldsymbol x_1,\boldsymbol x_2)  \left(\varphi(\boldsymbol x_1) - \varphi(\boldsymbol x_2)\right) \,dc(\boldsymbol x_1)\,dc(\boldsymbol x_2) =  \sum_{k=1}^{M_c}\frac{I_k}{S_{k}}\int_{\Omega_{C_{k}}}\varphi \,dxdy.
\end{align}
Above, the magnetic permeability $\mu$ is either a function of $\psi$ inside a region occupied by a ferromagnetic material, $\mu=\mu(|\nabla \psi|^2/x^2)$, or a constant 
$\mu = \mu_0$ elsewhere, $\boldsymbol x_i = (x_i,y_i)$ denote position vectors, $\Omega_{C_k}$ denotes the region occupied by the $k$-th coil, and the total number of coils is denoted by $M_c$. The following quantities, appearing on the left-hand side of \eqref{eq:VariationalFormulation}, are related to the Green's function associated with the operator on the left-hand side of \eqref{eq:FreeBoundarya} 
\begin{alignat*}{6}
N := \frac{1}{x} \left(\frac{1}{\delta_+} + \frac{1}{\delta_-} - \frac{1}{\rho}\right) &\, , \qquad
\delta_\pm := \sqrt{x^2 + (\rho \pm y)^2}\, ,
\qquad 
\kappa(\boldsymbol x_1,\boldsymbol x_2):= \sqrt{\frac{4x_1x_2}{(x_1 + x_2)^2 + (y_1-y_2)^2}}\, ,\\
M(\boldsymbol x_1,\boldsymbol x_2) :=\,&  \frac{\kappa(\boldsymbol x_1,\boldsymbol x_2)}{2\pi(x_1x_2)^{3/2}} \left(\frac{2-\kappa^2(\boldsymbol x_1,\boldsymbol x_2)}{2-2\kappa^2(\boldsymbol x_1,\boldsymbol x_2)}E(\kappa(\boldsymbol x_1,\boldsymbol x_2))- K(\kappa(\boldsymbol x_1,\boldsymbol x_2))\right)\,,
\end{alignat*}
where $E(\kappa(\boldsymbol x_1,\boldsymbol x_2))$ and $ K(\kappa(\boldsymbol x_1,\boldsymbol x_2))$ are complete elliptic integrals of the first and second kind respectively  \cite{Jackson:1975}.

\noindent\textit{Accounting for stochasticity.} We now consider the stochasticity in the currents and allow the vector of currents to be a $d$-dimensional random variable $\boldsymbol \omega$ uniformly distributed over the parameter space $W$ defined in \eqref{eq:ParameterSpace}. It is clear that in this case for any particular realization of the currents $\boldsymbol \omega$ we will obtain a \textit{different} equilibrium configuration $\psi(\boldsymbol \omega)$ that belongs to the Banach space $Z$ defined in \eqref{eq:FunctionSpace}. Moreover, since for every $\boldsymbol\omega\in W$ the resulting equilibrium configuration has finite energy, it then holds for the expected value of the equilibrium that
\[
\mathbb E\left(\|\psi\|_Z^2\right)<\infty.
\]
In mathematical terms, we say that the stochasticity of the currents transforms the solution $\psi$ to \eqref{eq:FreeBoundary} into a Banach space-valued random variable with finite expected energy. If $\boldsymbol \omega$ belongs to a complete and separable probability space $\left(W,\Sigma,\mathbb{P}\right)$, the class of such random variables forms what is known as a \textit{Bochner space} \cite{DaPrZa:2014}. In our particular case, solutions to \eqref{eq:FreeBoundary} are mappings from the parameter space $W$ to the Banach space $Z$ that, as functions of $\boldsymbol\omega$, belong to the Bochner space
\[
L^2(W,\Sigma,\mathbb{P};Z):=\{u:W\rightarrow Z\;\big\vert \; u \text{ strongly measurable,}\;\;\|u\|_{L^2(W,Z)}<\infty\},
\]
where the norm $\|\cdot\|_{L^2(W,Z)}$ is precisely defined in terms of the expected energy
\[
\|u\|_{L^2(W,Z)}  : = \left(\int_W \left\Vert u(\cdot,\omega) \right\Vert_{Z}^2 d\mathbb{P}(\omega) \right)^{1/2} = \left(\mathbb{E}\left(\left\Vert u(\cdot,\omega) \right\Vert_{Z}^2\right)\right)^{1/2}.
\]

\subsection{Mathematical results on multilevel Monte Carlo}
The proceeding analysis on the computational cost to compute the MLMC-FE estimator in terms of the desired relative accuracy $\epsilon$ is based on the following Theorem (established and proven in  \cite{ClGiScTe:2011}), which quantifies the distribution of the computational effort across discretization levels in terms of the relation between the decay rate of the variance of the correction terms, $V_\ell$, and the increase of the computational cost, $C_\ell$, as the mesh is refined. In particular, it states that if the variance of the correction terms decays faster than the increase of computational cost, the dominant computational expense takes place on the coarsest grid---see also \cite{Gi:2015}. 

\begin{theorem}
\label{thm:Theorem2}
Suppose there exist positive constants $a, b, c$ such that $a\ge \frac 1 2\min(b,c)$,
\begin{enumerate}
    \item [(i)] $\left\Vert\mathbb{E}\left(u-u_\ell\right)\right\Vert_Z=\mathcal{O}\left( M_\ell^{-a}\right)$,
    \item [(ii)] $V_\ell= \mathcal{O}\left(M_\ell^{-b}\right)$,
    \item [(iii)] $C_\ell=\mathcal{O}\left( M_\ell^{c}\right)$.
\end{enumerate}
Then for any  positive $\epsilon<e^{-1}$ small enough, there exists level $L$ and sample size $N_\ell$ for which the multilevel estimator $A_{\text{MLMC}}$ has an nMSE with 
\[
\frac{\mathbb E \left[\left\Vert\mathbb{E}(u)-A_{\text{MLMC}}(u_L) \right\Vert_{Z}^2\right]}{\mathbb E\left[\left\Vert\mathbb{E}(u) \right\Vert_{Z}^2\right]}<\epsilon^2,
\]
and the total computation cost $C$ with bound
\[
C\left(A_{\text{MLMC}} \right)=\left\{\begin{array}{ll}
\mathcal{O}\left(\epsilon^{-2}\right), & b>c,\\
\mathcal{O}\left(\epsilon^{-2}\left(\log \epsilon\right)^2\right), & b = c,\\
\mathcal{O}\left(\epsilon^{-2-\frac{(c - b)}{a}}\right), & 0<b<c.
\end{array}
\right.
\]
\end{theorem}

%
\bibliographystyle{abbrv}
\bibliography{references}
\end{document}